\documentclass[12pt,a4paper]{article}

\usepackage[T1]{fontenc}
\usepackage[utf8]{inputenc}
\usepackage[main=english,polish]{babel}
\usepackage[a4paper,margin=2.5cm]{geometry}
\usepackage{setspace}
\usepackage{newtxtext,newtxmath}
\usepackage{hyperref}
\usepackage{appendix}
\usepackage{float}
\usepackage{natbib}
\usepackage{graphicx}
\usepackage{graphicx,subcaption,lipsum}
\usepackage{caption}
\begin{document}

\begin{titlepage}
\setlength{\parindent}{0pt}

\vspace*{2cm}

\begin{center}
\LARGE\bfseries Application of Deep Reinforcement Learning to At-the-Money S\&P 500 Options Hedging
\end{center}

\vspace{1.5cm}

\begin{center}
\large
Zofia Bracha$^{1}$, Paweł Sakowski$^{2}$, Jakub Michańków$^{3}$\\[0.5cm]
\vspace{1.5cm}
\small
$^{1}$Faculty of Economic Sciences, University of Warsaw,\texttt{z.bracha@student.uw.edu.pl}\\
$^{2}$Department of Quantitative Finance and Machine Learning, University of Warsaw,\texttt{p.sakowski@uw.edu.pl}\\
$^{3}$TripleSun, Krakow, \texttt{jakub.michankow@triplesun.net}\\[0.5cm]

\end{center}

\vfill

\begin{center}
October 2025
\end{center}

\vspace{1cm}


\end{titlepage}

\section*{Abstract} 
\phantomsection
\addcontentsline{toc}{section}{Abstract} 

\vspace{1cm}

\setlength{\parindent}{1.5em}
\noindent
This paper explores the application of deep Q-learning to hedging at-the-money options on the S\&P 500 index. We develop an agent based on the Twin Delayed Deep Deterministic Policy Gradient (TD3) algorithm, trained to simulate hedging decisions without making explicit model assumptions on price dynamics. The agent was trained on historical intraday prices of S\&P 500 call options across years 2004 to 2024, using a single time series of six predictor variables: option price, underlying asset price, moneyness, time to maturity, realized volatility, and current hedge position. A walk-forward procedure was applied for training, which lead to nearly 17 years of out-of-sample evaluation. The performance of the deep reinforcement learning (DRL) agent is benchmarked against the Black–Scholes delta hedging strategy over the same time period. We assess both approaches using metrics such as annualized return, volatility, information ratio, and Sharpe ratio. To test models' adaptability, we performed simulations across varying market conditions and added constraints such as transaction costs and risk-awareness penalties. Our results show that the DRL agent can outperform traditional hedging methods, particularly in volatile or high-cost environments, highlighting its robustness and flexibility in practical trading contexts. While the agent consistently outperforms delta hedging, its performance deteriorates when the risk-awareness parameter is higher. We also observed that the longer the time interval used for volatility estimation, the more stable the results.
\vspace{2cm}
\begin{center}
\noindent\textbf{Keywords:}\\ Deep learning, Reinforcement learning, Double deep Q-networks, options market, options hedging, deep hedging\\
\vspace{2cm}
Thematic classification\\
C4, C14, C45, C53, C58, G13  
\end{center}
\clearpage
\tableofcontents
\newpage

\section{Introduction}
Hedging is a risk management technique, used to mitigate potential losses resulting from adverse price movements in assets~\citep{hull2018}. In the context of derivatives like options, hedging typically involves dynamically trading the underlying asset to offset the risk exposure introduced by holding the option. The most common approach is called delta hedging, which tries to offset the sensitivity of an option’s value to  movements in the underlying by maintaining a position proportional to the option’s delta. \par
In theory, this continuous rebalancing of the position eliminates all local risk associated with small price fluctuations of the underlying. However, the effectiveness of the hedge depends critically on the frequency of rebalancing, since in reality trading is discrete and subject to frictions, trading costs or liquidity constraints. Additionally, empirical asset returns often exhibit features such as jumps, heavy tails, and volatility clustering. In practice, the theoretical assumptions rarely hold and delta hedging in high cost or volatile environments can lead to losses. This has motivated the search for alternative hedging approaches that are better suited to real-world market dynamics.\par
In line with the growing interest in developing alternatives to classical methods, reinforcement learning (RL) approach has recently been introduced. Reinforcement learning is a machine learning technique that employs an agent to learn sequential decision-making under uncertainty. In RL, an agent interacts with environment by taking actions based on received feedback and then transferring to the next state, and
gradually learning a policy that maximizes expected return over the whole
period \citep{sutton2018}. In financial markets applications, this framing is natural: the agent corresponds to the trader, the environment to the market, and the reward function reflects a trade-off between risk reduction and transaction costs. Deep reinforcement learning extends this paradigm by employing deep neural networks to approximate value functions, enabling agents to handle high-dimensional and nonlinear state spaces that arise in realistic market settings. \par
Early work on applying deep reinforcement learning (DRL) to option hedging was pioneered by~\cite{buehler2019}. They introduced the Deep Hedging framework, demonstrating that deep neural networks can learn effective hedging policies in simulated markets, even when faced with frictions such as transaction costs, discrete rebalancing, and nonlinearities. Their results highlighted the flexibility of DRL methods compared to traditional stochastic control approaches, particularly when market imperfections complicate classical solutions. \par
Subsequent works have expanded this line of research, for instance by enriching the simulated environment with additional frictions and multiple hedging instruments or by incorporating utility-based objective in the model ~\citet{cao2021deep}. They analysed the returns  and stability of different functions. Additionally, ~\citet{francois2025} incorporated the full implied-volatility surface into the simulated market, demonstrating that DRL strategies can substantially outperform classical methods. \par
Despite these promising simulation based results, most implementations rely on carefully constructed environments or synthetic data, leading to questions about parameters selection and applicability in the real-world market. Addressing this gap,~\citet{mikkila2023empirical} presented one of the first empirical studies, training a DRL model directly on historical S\&P 500 index option data. In their paper, they extracted thousands of short time series, then trained the model on the segmented episodes and benchmarked it against classic delta hedging. Their findings indicate that deep hedging strategies can outperform traditional delta hedging, particularly when accounting for transaction costs and liquidity effects.\par
While research on deep hedging with empirically observed data remains relatively limited, this study seeks to extend the literature by exploring a novel approach. We construct a single, continuous time series of prices of at-the-money call options on S\&P 500 with maturity of approximately 30 days. Our goal is to demonstrate that a DRL agent can outperform classical delta hedging method, without a careful path selection process like in \citet{mikkila2023empirical} paper. Additionally, we aim to showcase the superiority of our model over the benchmark across different trading cost levels and risk aversion of the agent.\par
The research question is:
\begin{itemize}
    \item Can the DRL agent achieve superior performance relative to traditional delta hedging strategy across varying market conditions?
\end{itemize}
The paper is structured in a following way: section 1 provides an introduction, followed by a review of the relevant literature in section 2. Section 3 presents the theoretical background of the study, while section 4 outlines the methodology of the project. Section 5 reports the empirical results, and section 6 concludes.
\section{Literature review}
\subsection{Alternative hedging}
The literature on alternative hedging approaches is broad, reflecting the fact that perfect replication in the Black-Scholes framework is rarely feasible in practice. While this paper concentrates on recent deep reinforcement learning methods, it's important to recognise the abundance of alternatives to classical delta hedging. One prominent example is variance-optimal hedging \citet{schweizer1995}, which minimizes the mean-squared hedging error in incomplete markets. This approach provides a systematic way to construct hedges when perfect replication is impossible, and it has been widely applied in quadratic hedging theory. Closely related are utility-based approaches like \citet{pham2000}, in which the investor selects hedging strategies that maximize the expected utility of terminal wealth. By explicitly modeling preferences and risk aversion, these methods link hedging to portfolio optimization and provide a more economically grounded rationale for trading decisions.\par
Another important line of research is local risk minimization approach by \citet{follmer1986,follmer1991}, which focuses on minimizing conditional risk at each time step rather than in an overall expected sense. This makes it particularly relevant in discrete-time and incomplete markets, where global variance minimization may not capture the hedger’s real constraints. A further development of this idea is quantile hedging studied by \citet{follmer1999}, which recognizes frequent limitations of exact replication. Instead of insisting on sure replication, quantile hedging maximizes the probability of meeting the payoff subject to cost constraints, providing a natural trade-off between hedge effectiveness and affordability. \par
Beyond these rather classical approaches, the more modern frameworks treat hedging as a stochastic control problem. \citet{bertsimas2001}, applies dynamic programming techniques to derive optimal hedging rules. These models account for transaction costs, or other market frictions. Alternatively, risk measure based hedging approaches directly minimize coherent risk measures such as Conditional Value-at-Risk, thereby aligning hedging objectives with regulatory and risk management practice \citep{rockafellar2000}. 
\subsection{Machine learning in hedging}
Building on classical approaches, the use of machine learning models for hedging emerges as a natural next step in the research advancement. While classical methods rely on explicit probabilistic models, machine learning offers a data-driven way to create effective strategies directly from observed market dynamics. Early contributions of \citet{hutchinson1994} demonstrated that neural networks and non-parametric regressions were able to approximate option prices with high accuracy and generate hedging strategies that frequently outperformed the Black–Scholes delta. This work provided some of the first evidence that data-driven methods could capture non-linearities and market features that traditional parametric models miss.\par
More recently, \citet{ruf2022machine} offered a systematic perspective on applying machine learning to hedging, with a focus on interpretable and computationally efficient methods. In particular, regression techniques, support vector regression, and tree-based ensembles were highlighted as effective tools to model hedging errors in empirical option data. Their findings indicate that these machine learning methods can yield smaller mean-squared hedging errors than simple linear regression, while retaining greater transparency and robustness compared to deep neural networks. Together, these contributions illustrate the potential of simpler frameworks that do not involve building neural networks to enhance hedging beyond the classical delta-based approaches.\par
Recent research has extended these insights with a focus on robustness. \citet{wu2023robust} propose risk-aware hedging techniques that explicitly account for model misspecification and regime changes, showing that their method achieves smaller tail losses compared to variance-optimal hedges. Similarly, \citet{hirano2023learn} introduce a method to learn hedging rules directly from data without specifying the underlying price process. Their results suggest that this approach delivers more stable hedging errors across market conditions, particularly when compared against delta hedging under misspecified dynamics. In parallel, \citet{ruf2021neural} investigate the performance of an artificial neural network (ANN). They propose a HedgeNet and evaluate it against traditional delta hedging and linear regression benchmark. All tested strategies aimed to minimize the hedging error. While the outperformance of the proposed ANN over the benchmarks was significant, the difference between the deep learning approach and linear regression was relatively small. The authors suggest that this may be due to the ANN framework’s superior ability to capture leverage effects. 
\subsection{Deep hedging}
The term of \textit{Deep hedging} was introduced by \citet{buehler2019}, who proposed a novel framework for optimal hedging based on deep reinforcement learning. The introduced strategy aims to maximize the reward function under a convex risk measure. The generality of this framework enable wide range of objective functions. \citet{buehler2019} proposed agent implementation that uses a policy-gradient approach, which directly estimates the hedging policy as a function of state variables. The article clearly demonstrates the great potential of deep hedging and sets foundation for further research. He strengthened and further elaborated his idea in the following studies \citet{buehler2021deepdrift}, \citet{murray2022continuous}, \citet{buehler2022deep}.\par
Following this work, \citet{cao2021deep} extended the framework by investigating alternative objective functions and methods for calculating returns. They compared two perspectives: an accounting approach, in which the P\&L reflects the mark-to-market value of the portfolio, and a cash-flow approach, in which only realized inflows and outflows from trades are considered. Their experiments showed that the accounting approach delivers superior results, though they ultimately adopted a hybrid method that balances the two. In addition, they implemented an actor–critic algorithm with two critics: one estimating expected costs and the other estimating the squared value of costs. This design improved robustness by enabling the agent to control not only average performance but also its variability.\par
Several subsequent studies emphasize how objective design shapes hedging behavior. For instance, \citet{franccois2025difference} show that when risk measures insufficiently penalize adverse outcomes, the difference between deep- and delta hedged portfolios can resemble statistical arbitrage rather than pure risk reduction. Using more conservative risk measures such as CVaR eliminates this effect and produces genuine hedging policies. Along related lines, \citet{neagu2025} apply deep hedging under a GJR–GARCH(1,1) model and benchmark a range of reinforcement learning algorithms. They find that Monte Carlo policy-gradient methods consistently outperform value-based approaches in terms of hedging error, with policy-gradient methods being the only ones to surpass the Black–Scholes delta baseline.\par
Alternative methodological novelties have also appeared. \citet{halperin2020qlbs} sets the discrete time Black–Scholes-Merton model as a risk-adjusted Markov Decision Process and solve it with Q-learning. Unlike the policy-gradient methods common in deep hedging, their value-based RL approach learns an action–value function that jointly provides option prices and hedging strategies. In frictionless settings, QLBS mimics delta hedging, while remaining extensible to alternative objectives and frictions. As a continuation of this line, \citet{halperin2022machine}, develops further machine learning applications to pricing and hedging problems in discrete time. In parallel, \citet{ruf2020neural} and \citet{ruf2021neural} study related deep learning approaches and provide a comprehensive literature review of these methods for option pricing and hedging.\par
Other contributions extend the scope of deep hedging to more complex environments. For example, \citet{kolm2020dynamic} propose reinforcement learning based dynamic hedging policies that explicitly account for regime switching and microstructure effects. Their results suggest that regime aware hedging adapts more effectively to volatility shifts than static strategies, thereby reducing exposure during turbulent periods. In contrast to such simulator-based approaches, \citet{mikkila2023empirical} pursue a purely data-driven strategy by training hedging agents directly on high-frequency historical stock–option data. In this paper, the model was trained on a 5 trading day long period of intra-day prices. In this project the considered prices of call options on S\&P 500 index with maturity varying from 5 to 70 days. They show that empirical deep hedging achieves robust out-of-sample improvements relative to classical baselines, demonstrating the feasibility of real data training without specifying a volatility model. By comparison, sim-to-real frameworks such as \citet{francois2} have reported promising results, but remain restricted to training on simulated paths rather than full historical datasets.
\section{Theoretical background}
\subsection{Black-Scholes model (BSM) and hedging}
One of the most important model in the financial industry was introduced by Black and Scholes~\citep{black1973}, and independently extended by Merton in 1973~\citep{merton1973}. It assumes a market with constant volatility and interest rates, allowance of continuous trading without transaction costs or liquidity constraints. It is formulated as follows:
\begin{equation}
    dS_t = \mu S_t dt + \sigma S_t dW_t,
\end{equation}
where $\mu$ is the drift, $\sigma$ the volatility, and $W_t$ a standard Brownian motion, and $S_t$ a stock price. The arbitrage-free price of a European call option at time $t$ with strike $K$, maturity $T$, and risk-free rate $r$ is given by
\begin{equation}
    C(S_t, t) = S_t \Phi(d_1) - K e^{-r(T-t)} \Phi(d_2),
\end{equation}
where:
\begin{equation}
    d_1 = \frac{\ln\!\left(\frac{S_t}{K}\right) + \left(r + \tfrac{1}{2}\sigma^2\right)(T-t)}{\sigma \sqrt{T-t}},
\quad
d_2 = d_1 - \sigma \sqrt{T-t},
\end{equation}
and $\Phi(\cdot)$ denotes the standard normal cumulative distribution function.\par
An important derivation from the BSM model is option delta ($\Delta$), defined as the sensitivity of the option price with respect to changes in the underlying asset price:
\begin{equation}
    \Delta = \frac{\partial C}{\partial S_t}
\end{equation}
For European style call option, $\Delta = \Phi(d_1)$.\par
From this derivation, we can interpret delta as a hedge ratio that offsets the exposure of a short option position to small movements in $S_t$. In other words, this entails holding $-\Delta$ units of the underlying per option written. As the underlying asset price evolves over time, $\Delta$ changes accordingly, requiring the hedge position to be continuously rebalanced to preserve neutrality. Such strategy is called delta hedging and it is the most common approach to hedging.\par
In theory, this continuous rebalancing eliminates all local risk associated with small price fluctuations of the underlying. The effectiveness of the hedge, however, depends critically on the frequency of rebalancing, since in reality trading is discrete and subject to frictions. The underlying model was introduced  Under these idealized conditions, the model delivers a theoretically perfect hedging strategy through continuous delta rebalancing, such that a riskless portfolio can be constructed and replicated by dynamically adjusting exposure to the underlying asset.
\subsection{Volatility}
Volatility, defined as the magnitude of fluctuations in the price of a financial asset, plays a central role in asset pricing, risk management, and portfolio allocation. Since it's an unobservable quantity, numerous methodologies have been developed to estimate and forecast volatility. 
The three main methods of estimating volatility are: realized (historical) volatility, implied volatility and stochastic volatility.
\subsubsection{Realized volatility}
It is usually calculated as a statistical measure of past fluctuations in asset returns, most commonly as the standard deviation of log-returns over a specified window. The advantage of using this method is its simplicity in both understanding as well as implementation. However, realized volatility is calculated using past data and it may be inaccurate in reflecting sudden changes in current market conditions. \citet{andersen2003modeling}
\subsubsection{Implied volatility}
In contrast, implied volatility is a measure considering future outlook. It is calculated from the prices of traded options under an assumed pricing model such as BSM model. It represents the market’s expectation of future variability, embedding both the statistical expectation of variance and a risk premium demanded by option sellers. Because implied volatility reflects current market sentiment and expectations, it is often more responsive to new information than historical measures. However, its estimation is inherently model-dependent and may vary across strikes and maturities, leading to volatility smiles and surfaces.
\subsubsection{Stochastic volatility}
In this framework, volatility is modeled as a stochastic process, meaning it is treated as a latent random variable that evolves over time. Prominent examples include the GARCH family of models \citep{engle1982}, \citep{bollerslev1986}, the Heston model \citep{heston1993}, and the SABR \citep{hagan2002} model, each of which captures different aspects of volatility dynamics such as persistence, mean reversion, or stochastic skew. A key challenge in their practical implementation is the calibration of model parameters, as accurate estimation is crucial for ensuring that the model reliability.\par
Over the years, multiple studies investigated efficiencies of all of these types in different settings (\citet{christensen1998}, \citet{ammann2009}, \citet{sjöberg2023}). In deep hedging, different approaches were used to estiamte volatility inlcuding implied volatility or simulated with deifferent models. To evaluate a more simplistic approach, in this paper we have decided to use realized volatility to estimate hedge ratio.
\subsection{Reinforcement Learning}
Reinforcement Learning is a machine learning technique in which an agent interacts with an environment over a sequence of discrete time steps ~\citep{sutton2018}. At the core of RL is the concept of learning through trial and error, guided by a reward signal. In other words, the model tries to maximise a defined cumulative reward objective function $G_t$ over the time period T. At each time step $t \in \{0,1,2,\ldots\}$, the agent observes the current state $s_t \in \mathcal{S}$, selects an action $a_t \in \mathcal{A}$, and receives a scalar reward $r_t \in \mathbb{R}$. The environment then transitions to a new state $s_{t+1}$ according to its dynamics $P(s’ \mid s,a)$\par
The goal of the agent is to learn a policy $\pi$, a function mapping states to actions, that maximizes the expected cumulative return. As introduced in \citet{sutton2018}, the cumulative return $G_t$ is typically defined as the discounted sum of future rewards:
\begin{equation}
    G_t = \sum_{k=0}^\infty \gamma^k r_{t+k+1}.
\end{equation}
\subsubsection{Markov Decision Process (MDP)}
\begin{figure}[h]
\centering
\includegraphics[width=0.8\textwidth]{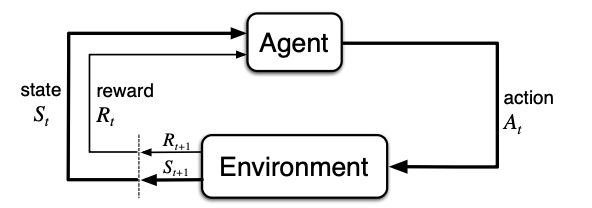}
\caption{Interactions between objects in RL from~\citet{sutton2018}}
\label{fig:RL}
\end{figure}

The formal model underlying reinforcement learning is the Markov decision process, defined as a tuple of 4 variables $\mathcal{M} = (\mathcal{S}, \mathcal{A}, P, R)$,
where each element is defined as follows:
\begin{itemize}
    \item \textbf{State space} $\mathcal{S}$: the set of all possible states
    \item \textbf{Action space} $\mathcal{A}$: the set of actions available to the agent
    \item \textbf{Transition kernel $P(s’|s,a)$}: the probability of transitioning to state $s’ \in \mathcal{S}$ when action $a \in \mathcal{A}$ is taken in state s
    \item \textbf{Reward function R(s,a)}: the expected reward obtained at time t immediately after taking action a at state s
\end{itemize}
Figure~\ref{fig:RL} below from \citet{sutton2018} illustrates the interactions between different variables.\\
An MDP satisfies the Markov property that states that next state depends only on the current state and action:
\begin{equation}
\Pr(s_{t+1} | s_0, a_0, s_1, a_1, \ldots, s_t, a_t) = \Pr(s_{t+1} | s_t, a_t)
\end{equation}
To evaluate policies and determine actions, we can define value functions that quantify the expected return:
\begin{itemize}
    \item The state-value function $ V^\pi(s)$  denotes the expected return starting from state s and following a policy $\pi$:
    \begin{equation}
        V^\pi(s) = \mathbb{E}_\pi \left[ G_t \,\middle|\, s_t = s \right]
    \end{equation}
    \item The action-value function $Q^\pi(s, a)$ represents the expected return from state s, when taking action a, and then following policy $\pi$:
    \begin{equation}
        Q^\pi(s,a) = \mathbb{E}_\pi \left[ G_t \,\middle|\, s_t = s,\, a_t = a \right]
    \end{equation}
\end{itemize}
The optimal values for these functions are then:
\begin{equation}
V(s) = \max_\pi V^\pi(s), \quad Q(s,a) = \max_\pi Q^\pi(s,a).
\end{equation}
And the optimal policy $\pi(s)$ is:
\begin{equation}
    \pi^* (s) = \arg\max_{a \in \mathcal{A}} Q^*(s,a)
\end{equation}
\subsubsection{Bellman equations}
Another key puzzle in the reinforcement learning frameworks are Bellman equations, which express a recursive relationship between the two value functions~\citet{bellman1952}. For any policy $\pi$, the Bellman equations are:
\begin{equation}
    V^\pi(s) = \sum_{a \in \mathcal{A}} \pi(a|s) \Big[ R(s,a) + \gamma \sum_{s’ \in \mathcal{S}} P(s’|s,a) V^\pi(s’) \Big],
\end{equation}
\begin{equation}
    Q^\pi(s,a) = R(s,a) + \gamma \sum_{s’ \in \mathcal{S}} P(s’|s,a) \sum_{a’ \in \mathcal{A}} \pi(a’|s’) Q^\pi(s’,a’)
\end{equation}\\
And the Bellman optimality equations are:
\begin{equation}
V^* (s) = \max_{a \in \mathcal{A}} \Big[ R(s,a) + \gamma \sum_{s’ \in \mathcal{S}} P(s’|s,a) V^(s’) \Big],
\end{equation}
\begin{equation}
Q^* (s,a) = R(s,a) + \gamma \sum_{s’ \in \mathcal{S}} P(s’|s,a) \max_{a’ \in \mathcal{A}} Q^(s’,a’)
\end{equation}
\subsubsection{Temporal Difference Learning}
Temporal difference learning is a foundational approach to estimate value functions that doesn't require complete model of the environment to make predictions. Unlike Monte Carlo methods, which require complete episodes to update value estimates, TD learning updates its predictions incrementally at each time step using bootstrapping \citet{sutton1988}. TD learning underpins widely used algorithms such as TD(0), SARSA, and Q-learning, and is particularly well-suited for financial modelling.\\
The simplest form, TD(0), updates the state-value function as follows:
\begin{equation}
   V(s_t) \leftarrow V(s_t) + \alpha \Big[ r_t + \gamma V(s_{t+1}) - V(s_t) \Big], 
\end{equation}
where $\alpha$ is the learning rate. The term in brackets is the TD error
\begin{equation}
    \delta_t = r_t + \gamma V(s_{t+1}) - V(s_t),
\end{equation}
which quantifies the difference between predicted and observed returns.
\subsubsection{Q-learning}
Introduced by \citet{watkins1992}, Q-learning is an off-policy temporal difference learning framework, which directly estimates the optimal action-value function $Q^*(s,a)$:
\begin{equation}
Q(s_t, a_t) \leftarrow Q(s_t, a_t) + \alpha \Big[ r_t + \gamma \max_{a'} Q(s_{t+1}, a') - Q(s_t, a_t) \Big]
\end{equation}
In reinforcement learning, off-policy means that the agent learns about the target policy $\pi$, while following a different behavior policy $\beta$ to generate experience. While the methods above provide strong promises, in practice they suffer from scalability issues in environments with large or continuous state and action spaces. To address this, reinforcement learning has been combined with deep learning techniques, giving rise to deep reinforcement learning, where neural networks approximate policies and value functions.
\subsection{Deep Reinforcement learning}
Classical reinforcement learning methods such as dynamic programming or tabular temporal-difference learning rely on explicit representations of value functions and policies. While effective in small state and action spaces, these methods become intractable in high dimensional or continuous domains due to the curse of dimensionality. To overcome this limitation, reinforcement learning has been combined with deep learning techniques, resulting in the field of Deep Reinforcement Learning \citep{mnih2015}.\par
Deep RL leverages deep neural networks as function approximators for the value function, action-value function, or policy. Instead of storing values for every state-action pair in a table, a parameterized neural network $Q(s,a;\theta)$ or $\pi(a|s;\theta)$ is trained, where $\theta$ denotes the network parameters. This enables generalization across large state spaces and allows learning in complex environments such as video games, robotics, or financial markets.

\subsubsection{Deep Q-Networks (DQN)}
One of the first major model combining deep learning and reinforcment learning was the Deep Q-Network (DQN), introduced by \citet{mnih2015}. DQN approximates the optimal action-value function $Q^*(s,a)$ using a deep neural network. The update rule extends the classical Q-learning update:
\begin{equation}
    \theta \leftarrow \theta + \alpha \Big[ y_t - Q(s_t,a_t;\theta) \Big] \nabla_\theta Q(s_t,a_t;\theta),
\end{equation}
where the target $y_t$ is defined as:
\begin{equation}
    y_t = r_t + \gamma \max_{a'} Q(s_{t+1},a';\theta^-).
\end{equation}
One drawback of DQN is the systematic overestimation of action values due to the maximization operator applied to noisy value estimates. To mitigate this bias,~\cite{vanhasselt2015} proposed double DQN that separates action selection and evaluation and reduces bias and improve stability:
\begin{equation}
    y_t = r_{t+1} + \gamma \, Q\!\left(s_{t+1}, \arg\max_{a'} Q(s_{t+1},a';\theta)\,;\,\theta^-\right)
\end{equation}
\subsubsection{Deep Deterministic Policy Gradient (DDPG)}
While double DQN are effective in discrete action spaces, computing $\max_{a'} Q(s',a')$ is impossible in continuous setting, hence the motivation for using policy-based and actor-critic methods. The Deep Deterministic Policy Gradient (DDPG) algorithm, introduced by~\citet{lillicrap2016}, extends the deterministic policy gradient framework by leveraging deep neural networks and key stabilizing techniques from DQN.
DDPG employs two neural networks:
\begin{itemize}
    \item \textbf{Actor network} $\mu_\theta(s)$: outputs a deterministic action given state $s$.
    \item \textbf{Critic network} $Q_\phi(s,a)$: approximates the action-value function for the current policy.
\end{itemize}
To improve stability, DDPG uses:
\begin{itemize}
    \item \textbf{Target networks} $\mu_{\theta'}$ and $Q_{\phi'}$, which are slowly updated copies of the actor and critic used to compute stable targets.
    \item \textbf{Experience replay buffer} $\mathcal{D}$, from which minibatches of past transitions $(s,a,r,s')$ are sampled to break temporal correlations.
\end{itemize}
The critic parameters $\phi$ are updated by minimizing the TD error:
\begin{equation}
    L(\phi) = \mathbb{E}_{(s,a,r,s') \sim \mathcal{D}} \left[ \Big( Q_\phi(s,a) - y \Big)^2 \right],
\end{equation}
with target:
\begin{equation}
    y = r + \gamma Q_{\phi'}(s', \mu_{\theta'}(s')).
\end{equation}
The actor is updated using the deterministic policy gradient:
\begin{equation}
    \nabla_\theta J(\theta) \approx \mathbb{E}_{s \sim \mathcal{D}} \Big[ \nabla_\theta \mu_\theta(s) \, \nabla_a Q_\phi(s,a) \big|_{a = \mu_\theta(s)} \Big].
\end{equation}
Although DDPG was a major step forward in continuous control, it suffers from several issues:
\begin{itemize}
    \item \textbf{Overestimation bias} in the critic due to using a single Q-function.
    \item \textbf{High variance and instability} in learning when the critic is poorly estimated.
    \item \textbf{Sensitivity to hyperparameters} such as learning rates, noise scale, and target update rate.
\end{itemize}
These shortcomings motivated the development of more robust actor–critic algorithms, such as Twin Delayed Deep Deterministic Policy Gradient (TD3).
\subsubsection{Twin Delayed Deep Deterministic Policy Gradient (TD3)}
TD3 introduced by \citet{fujimoto2018} represents a significant improvement in stability and performance over earlier actor–critic methods. The three main modifications made to DDPG to address its limitations are:
\begin{itemize}
    \item \textbf{Double Q-learning clipping}: two critics, $Q_{\phi_1}$ and $Q_{\phi_2}$, are trained in parallel. The target is computed using the minimum of their estimates to reduce overestimation bias:
    \begin{equation}
    y_t = r_{t+1} + \gamma \min_{i=1,2} Q_{\phi_i’}(s_{t+1}, \mu_{\theta’}(s_{t+1}) + \epsilon),
    \end{equation}
    where $\epsilon$ is clipped Gaussian noise
    \item \textbf{Target policy smoothing}: the action in the target is perturbed by $\epsilon \sim \mathcal{N}(0, \sigma^2)$, clipped to a range [-c, c]. This prevents the critic from overfitting. The bounds in this project are [-1,1]
    \item \textbf{Delayed policy updates}: the actor and target networks are updated less frequently, reducing the likelihood that the policy is updated based on untrained critics.
\end{itemize}
TD3 retains the off-policy nature of DDPG, using experience replay and noisy behavior policies for exploration, but vastly improves robustness and sample efficiency. These properties make TD3 especially suitable in environments where robustness to noisy estimates and stability of training are critical.
\section{Methodology}
\subsection{Problem formulation}
In this project we model dynamic option hedging as a finite‑horizon Markov Decision Process Markov decision process $\mathcal{M}=(\mathcal{S},\mathcal{A},P,R)$ indexed by trading times $t=1,\dots,T$ aligned with a historical price path. The agent holds a short option position and trades the underlying to hedge. At time t, the state $s_t$ summarizes market observed variables and the current hedge; the action $a_t\in\mathbb{R}$ adjusts the hedge position; and the reward $r_t$ is the risk‑adjusted incremental hedging PnL. We define the problem as follows:
\begin{itemize}
    \item \textbf{State space}: 6-dimensional observation vector at each step: option price, underlying price, time-to-maturity, moneyness, realized volatility, current hedge position. These features are predictor variables for the DRL model.
    \item \textbf{Action space}: actions are one-dimensional and represent a new hedge ratio. They are clipped to [-1, 1] and mapped to a bounded change in position.
    \item \textbf{Dynamics}: Transitions are driven by the supplied market time series and the internal portfolio recursion. The environment is episodic and advances deterministically through timestamps
    \item \textbf{Reward}: Risk-adjusted cumulative reward (PnL) described in details later.
\end{itemize}
\subsubsection{Objective function}
One of the most important step is the objective function selection. There have been multiple approaches to this topic. \citet{cao2021deep} discussed different PnL calculation techniques for objective function. \citet{francois2025} used a different approach and included risk measures in the objective function. \citet{mikkila2023empirical} discussed using variance or standard deviation of temporal rewards. Based on this, we have decided to define the maximisation problem as follows:
\begin{equation}
    \max \mathbb{E}[w_T] - \xi  SD(w_T),
\end{equation}
which is estimated at each step as:
\begin{equation}
    R_t = PnL_t - \xi |PnL_t|,
\end{equation}
where $\xi$ denotes the linear risk penalty parameter applied to discourage extreme outcomes. This general approach of maximizing risk-adjusted profit originates from the mean–variance utility framework introduced by \citet{markowitz1952} and further developed in \citet{markowitz1959}. \par
The $PnL_t$ is calculated as:
\begin{equation}
    \text{PnL}_t = HO_t \cdot (C_t - C_{t-1}) + HS_{t-1} \cdot (S_t - S_{t-1}) - c \cdot |S_t| \cdot |\Delta HS_t|
\label{eq:pnl}
\end{equation}
where c represents transaction costs, HS holding of stock, S stock price, HO holding of option, and finally C represents option price.
\subsection{Model implementation}
In this paper we used Twin Delayed Deep Deterministic Policy Gradient (TD3) framework that was introduced in the previous section. As a benchmark model we use traditional delta hedging model, which means that at each time step we adjust our hedge position based on the current prices. Across all periods we assumed 2\% risk free rate. The model was implemented using Pytorch library \citet{paszke2019pytorch}. While the description below should present all details of the implementation, we include a link to the project Github repository with more details at the end of the article.
In this project, similar networks structure was used to \citet{mikkila2023empirical}. The actor network is a deterministic feed-forward function mapping states to continuous actions. It consists of two hidden layers of size 256 with LeakyReLU activations (negative slope 0.05) and a final Tanh output that ensures the raw action lies within (-1,1).\par
Both critic networks approximate the state-action value function. They take as input both the state vector and action and process them through three hidden layers of size 256, each followed by LeakyReLU activation function with slope parameter 0.05. The networks terminate in a single linear output representing the scalar Q-value.\par
The Mean Squared Error (MSE) loss function is used to assess the quality of the critics’ Q-value approximation, while the Adam optimizer updates the actor and critic network weights via gradient descent to reduce this loss and improve the model’s predictive power.
\subsubsection{Exploration noise}
As highlighted earlier, in models like TD3 exploration must be introduced explicitly because the policy itself is deterministic. Noise is a small stochastic incremanet added to the action that was predicted by the model. In the present framework, two complementary sources of noise were used. The primary mechanism is an Ornstein–Uhlenbeck (OU) process introduced by \citet{ornstein1930brownian} applied directly to the agent’s actions. Formally, the Ornstein–Uhlenbeck is defined as:
\begin{equation}
    dX_t = \theta (\mu - X_t) dt + \sigma dW_t
\end{equation}
where $X_t$ denotes the noise state, $\mu$ the long-term mean, $\theta$ the rate of mean reversion, $\sigma$ the volatility parameter, and $W_t$ a standard Wiener process.
This process generates smooth and continuous trajectories, thus ensuring that successive actions remain temporally correlated with mean-reverting properties.
Additionally, the added noise decreases over time. Early in training, the larger perturbations encourage broad exploration of the action space, while in later stages the reduced scale allows the agent to stabilize its strategy. The OU noise is scaled by current parameter $a_t$, which moves form initial high noise $a_0$ to $a_{min}$ defined as follows:
\begin{equation}
    \alpha_t = \alpha_0 - (\alpha_0 - \alpha_{\min}) \cdot \min\left(1, \frac{t}{N_{\text{decay}}}\right).
\end{equation}
Similar noise approach was used in \citet{lillicrap2016}.\par
The second noise mechanism operates within the TD3 training procedure itself, when the algorithm adds a small Gaussian perturbation to the target policy’s action while computing Bellman targets for the critics. Such noise is also clipped to lie within defined bounds (here [-1,1]). This policy smoothing prevents the critics from overfitting to sharp peaks in their value estimates, thereby reducing overestimation bias and encouraging more conservative learning. Together, these two sources of noise at action selection and smoothing noise in target evaluation provide both good exploration and robustness against critic instability.
\subsubsection{Replay buffer}
To complement exploration, the agent employs a replay buffer, which is central to off-policy reinforcement learning. The buffer stores past interactions with the environment as a tuple of transitions $(s_t, a_t, r_t, s_{t+1})$, where $s_t$ is the state, $a_t$ the action taken, $r_t$ the resulting reward, $s_{t+1}$ the next state. The buffer has defined capacity (10000) and operates under a first-in–first-out replacement scheme, so that older experiences are gradually replaced by more recent ones. During training, minibatches of fixed size are drawn uniformly at random from the buffer. This procedure breaks the strong temporal correlations that naturally exist in financial time series, producing approximately i.i.d. samples that are better suited for stochastic gradient descent updates. The replay buffer prevents the critic networks from overfitting to spikes, preserves exploratory actions from early training and prevents harmful feedback loops that could lead to actor overfitting. 
\subsubsection{Volatility estimation}
In this paper we employ a realized volatility framework . First, asset prices $P_t$ are transformed into log-returns, defined as
$r_t = \ln\left(\frac{P_t}{P_{t-1}}\right)$.
Then the volatility is computed using a rolling standard deviation of log-returns over a specified window length of $w$ data points. Formally, the realized volatility at time $t$ is:
\begin{equation}
    \sigma_t = \sqrt{A} \cdot SD\left(r_{t-w+1}, \ldots, r_t\right),
\end{equation}
where $A$ is an annualization factor to scale volatility estimates to an annualized level.\\
At the beginning of a time period, when there is less data points than w, the volatility is calculated on rolling basis. Since the volatility is calculated using returns, the first volatility value is impossible to be calculated hence it is replaced with a randomly drawn volatility from a uniform distribution from range [0.02, 0.10]. Volatility window is one of the parameters, on which we perform sensitivity analysis.
\subsection{Data}
For the project we used historical prices of at-the-money call options on S\&P 500 with 30 day maturity and S\&P 500 itself from years 2004-2024. The data was sourced from CBOE DataShop. For the purpose of this project we have decided to use 30 min frequency for rebalancing. For each time step, the most representative option contract was selected that had the smallest deviations from 30-day maturity and at-the-money moneyness ($moneyness \approx 1$), thus ensuring consistency in option characteristics. This selection criterion mitigates noise and structural shifts due to varying contract availability. To address potential anomalies and discontinuities in price series, a forward-filling mechanism was applied to replace invalid or zero values with the first valid observation, followed by a price-capping procedure that constrained daily price movements within a ±20 \% band relative to the previous value. This capping method reduces the influence of outliers that are more likely to be an input error rather than a real observation. Additionally, before training the models, all input data was normalized using Z-score normalisation to improve training.
\subsection{Training}
\subsubsection{Walk forward approach}
The walk-forward approach is a common technique for training and evaluating machine learning models on time series data \citep{slepaczuk}. Unlike conventional data splitting methods, which partition the available dataset into fixed and disjoint training, validation, and testing subsets, the walk forward method relies on an iterative scheme that more closely reflects the nature of real-world forecasting tasks.\par
\begin{figure}[h]
\centering
\includegraphics[width=0.8\textwidth]{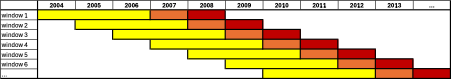}
\caption{Ilustration of a walk forward approach produced using Excel}
\label{fig:walk}
\end{figure}

In this framework, the model is trained on an initial window of historical data and subsequently validated on the immediately following time interval. After evaluation, the training window is rolled forward to include this validation segment, and the process is repeated (Fig.~\ref{fig:walk}). In each iteration, the model is re-estimated using the most recent information, and performance is assessed on the next unseen period. This procedure continues until the end of the dataset is reached, resulting in a sequence of out-of-sample forecasts that collectively provide a robust estimate of the model’s predictive power.\par
Such approach offers several advantages:
\begin{itemize}
    \item Ensures no future information is used to predict past outcomes (temporal leakage risk).
    \item Allows the model to continuously adapt to evolving data-generating processes, which is particularly valuable in non-stationary environments where structural breaks, regime shifts, or gradual drifts may occur.
    \item Due to combining multiple out-of-sample periods, the method reduces the variance associated with performance evaluation compared to a single fixed test set, giving more reliable insights into model generalizability.
\end{itemize}
\subsubsection{Hyperperameter tunning}
Within each window, hyperparameter tunning is conducted to ensure the model parameters are well adjusted. The hyperparameters that were tunned are presented in the table ~\ref{tab:hyperparams} and include exploration parameters, TD3 parameters, and others. reasonable ranges of parameters for tunning were selected during initial tests performed on first window (years 2004-2008). In this project we employ the hyperparameter tunning in each window. Hyperparameters are tuned exclusively on the training and validation time periods to avoid information leakage, with the validation year performance serving as the criterion for selecting the best configuration. To explore the hyperparameter space efficiently, we implement a randomized search procedure in which five candidate parameter sets are drawn from the predefined ranges provided in Table~\ref{tab:hyperparams}. The best-performing configuration, as determined by validation performance, is then retained and evaluated on the subsequent test year. This process is repeated at each walk-forward step, ensuring that the model is continuously adjusted in response to new data.
\begin{table}[H]
\centering
\begin{tabular}{l c}
\hline
\textbf{Parameter} & \textbf{Range} \\
\hline
\multicolumn{2}{l}{\textbf{TD3 Algorithm}} \\
Soft update coefficient ($\tau$) & $(0.001, 0.01)$ \\
Discount factor ($\gamma$) & $(0.95, 0.999)$ \\
Policy noise & $(0.1, 0.3)$ \\
Noise clip & $(0.2, 0.5)$ \\
Policy update frequency & $(1, 5)$ \\
\multicolumn{2}{l}{\textbf{Learning Rates}} \\
Actor learning rate & $(10^{-7}, 10^{-3})$ \\
Critic learning rate & $(10^{-7}, 10^{-3})$ \\
Batch size & $(64, 512)$ \\
\hline
\multicolumn{2}{l}{\textbf{Exploration}} \\
Initial noise & $(0.2, 0.5)$ \\
Final noise & $(0.01, 0.1)$ \\
Noise decay steps & $(50{,}000, 200{,}000)$ \\
Ornstein–Uhlenbeck $\theta$ & $(0.1, 0.3)$ \\
Ornstein–Uhlenbeck $\sigma$ & $(0.1, 0.3)$ \\
\hline
\multicolumn{2}{l}{\textbf{Environment}} \\
Initial volatility & $(0.10, 0.40)$ \\
Minimum initial volatility & $(0.01, 0.07)$ \\
Maximum initial volatility & $(0.10, 0.40)$ \\
\hline
\end{tabular}
\caption{Hyperparameter ranges}
\label{tab:hyperparams}
\end{table}
\subsection{Performance metrics}
The performance of the model is measured in multiple ways. We calculate cumulative PnL as described in equation \eqref{eq:pnl} combined across all out-of-sample periods, normalized by notional=1000\$, so the value of 100 represents 100\% return on the notional. This approach enables comparison across different option strategies, time periods, and market conditions while maintaining computational stability. In the analysis, we do not account for margin requirements in the case of negative returns. As a result, cumulative return values may reach levels that would be unrealistic in a real-world hedging. Additionally, we compute following performance measures:
\subsubsection{Maximum drawdown (MDD)}
The maximum drawdown measures the largest observed loss from a peak to a trough of a portfolio’s value before a new peak is attained. It quantifies downside risk over a specified time horizon.
\begin{equation}
    \text{MDD} = \max_{t \in [0,T]} \left( \frac{\max_{\tau \in [0,t]} V(\tau) - V(t)}{\max_{\tau \in [0,t]} V(\tau)} \right)
\end{equation}
where V(t) is the portfolio value at time t, and T is the investment horizon.
\subsubsection{Annualized rate of return (ARC)}
The annualized rate of return is the geometric average return per year over the investment horizon.
\begin{equation}
    \text{ARC} = \left( \frac{V(T)}{V(0)} \right)^{\tfrac{1}{Y}} - 1
\end{equation}
\subsubsection{Annualized Standard Deviation (ASD)}
The annualized standard deviation measures the volatility of returns on an annual basis. If returns are observed at frequency m, the scaling is applied as:
\begin{equation}
    \text{ASD} = \sigma_r \cdot \sqrt{m}
\end{equation}
where:
\begin{itemize}
	\item	$\sigma_r$ is the standard deviation of periodic returns,
	\item	m is the number of periods per year.
\end{itemize}
\subsubsection{Information ratio}\
The Information Ratio measures the consistency of active returns relative to a benchmark by dividing average active return by the tracking error. 
\begin{equation}
    \text{IR} = \frac{E[R_p - R_b]}{\sigma(R_p - R_b)}
\end{equation}
where:
\begin{itemize}
	\item $R_p$ is the portfolio return,
	\item $R_b$ is the benchmark return,
	\item $\sigma(R_p - R_b)$ is the standard deviation of active returns.
\end{itemize}
Delta hedging strategy performance was used as benchmark in Information ratios calculations.
\subsubsection{Adjusted information ratio}
The Adjusted Information Ratio (AIR) used here is a custom performance metric that modifies the traditional Information Ratio by explicitly penalizing returns for risk through both volatility and drawdowns. It is defined as:
\begin{equation}
AIR = IR \cdot \frac{ARC^2 - \operatorname{sign}(ARC)}{ASD \cdot |MD|}
\end{equation}
where:
\begin{itemize}
\item $ARC$ is the Annualized Return Compounded,
\item $ASD$ is the Annualized Standard Deviation of returns,
\item $MD$ is the maximum drawdown,
\item IR is information ratio
\end{itemize}
\subsubsection{Sharpe ratio (SR)}
The Sharpe Ratio was introduced by \citet{sharpe1966mutual}. It evaluates the excess return per unit of total risk, where risk is measured by the standard deviation of portfolio returns.
\begin{equation}
    \text{SR} = \frac{E[R_p - R_f]}{\sigma(R_p - R_f)}
\end{equation}
where:
\begin{itemize}
    \item $R_p$ is the portfolio return,
	\item $R_f$ is the risk-free rate,
	\item $\sigma(R_p - R_f)$ is the standard deviation of excess returns.
\end{itemize}
\section{Results}
\subsection{Base case}
\begin{figure}[H]
\centering
\includegraphics[width=0.8\textwidth]{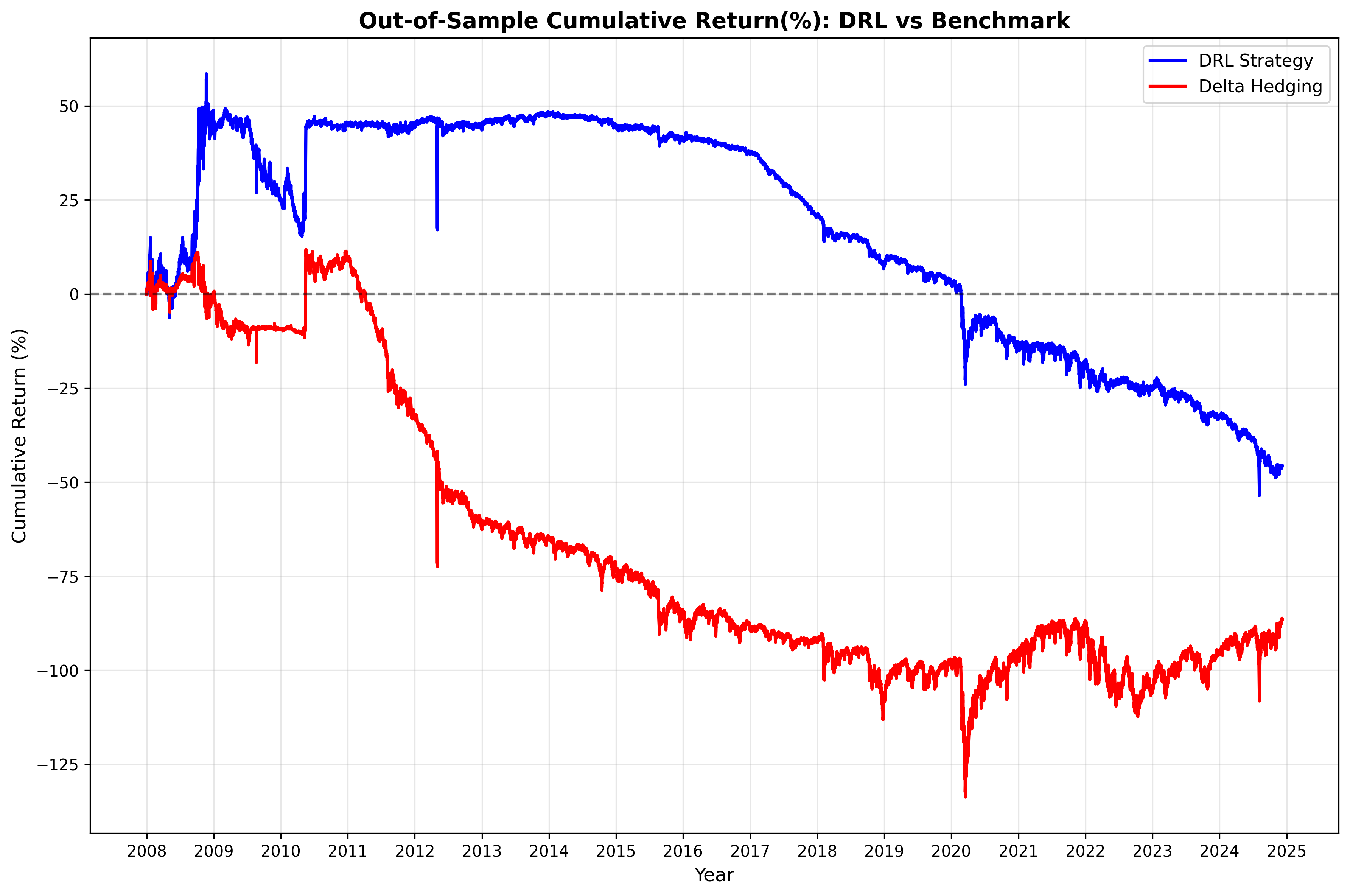}
\caption{Cumulative return}
\label{fig:cum_pnl}
\end{figure}

At first we run the comparison with trading cost (c) set to 0.1\%, risk penalty parameter ($\xi$) 1\%, and volatility computed using 50 data steps. We performed hyperparameter tuning with 5 random combinations of parameters from defined ranges. Figure \ref{fig:cum_pnl} below represents the cumulative return calculated for each out-of-sample period and combined together in one chart. The statistics included in the tables contain averages of the metrics calculated during each out-of-sample period. Results at Figure \ref{fig:cum_pnl} show that DRL model outperformed the benchmark, even though the return is negative. Additionally, we can observe worse performance of the model in years 2017-2024. We believe that in order for the DRL agent to adapt to the change in market dynamics, more extensive hyperparameter search could be required. 
\vspace{0.1cm}
The results in the Table \ref{tab:drl_vs_delta_setting2} show superior performance of the DRL agent over the classical delta hedging benchmark model. The DRL managed to get positive Information ratio, despite heavy drawdowns and negative returns.
\begin{table}[H]
\centering
\caption{Comparison of DRL vs. Benchmark in base case scenario}
\begin{tabular}{lcc}
\hline
{Metric} & {DRL} & {Benchmark (DH)} \\
\hline
Annualized return (ARC) & -0.40\% & -0.63\% \\
Annualized Std. dev. (ASD) & 0.0576 & 0.0629 \\
Sharpe Ratio & -0.417 & -0.418 \\
Information Ratio & 0.0325 & --- \\
Adj. Information Ratio & 23.06 & --- \\
Max Drawdown & -75.26\% & -81.90\% \\
Final PnL (\%) & -45.51\% & -85.64\% \\
\hline
\end{tabular}
\label{tab:drl_vs_delta_setting2}
\end{table}
\subsection{Transaction costs}
\begin{figure}[H]
\centering
\includegraphics[width=0.8\textwidth]{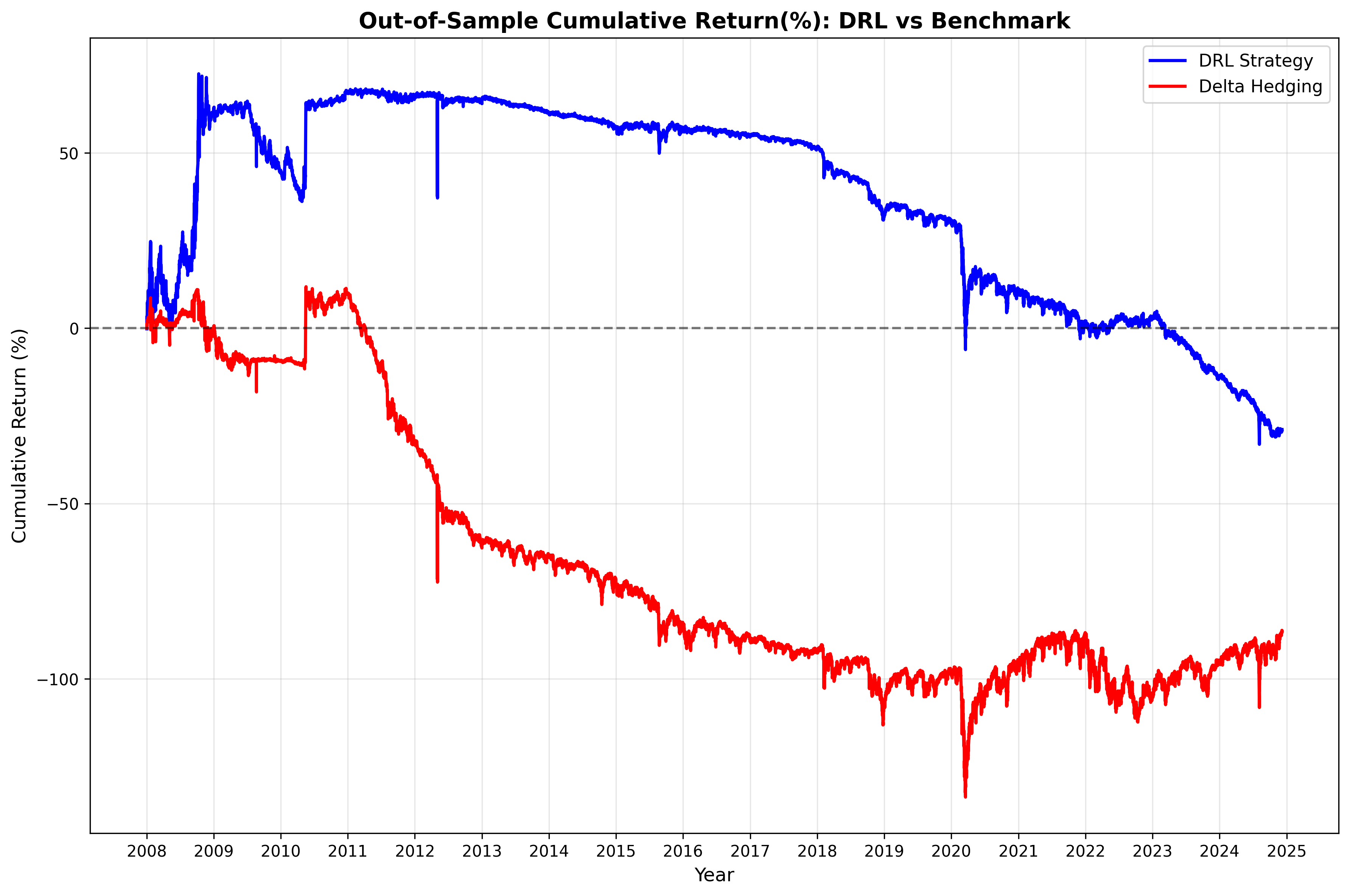}
\caption{Cumulative Return for $C_1$=0.1\%}
\label{fig:cum_pnl_bps10}
\end{figure}

\begin{figure}[H]
\centering
\includegraphics[width=0.8\textwidth]{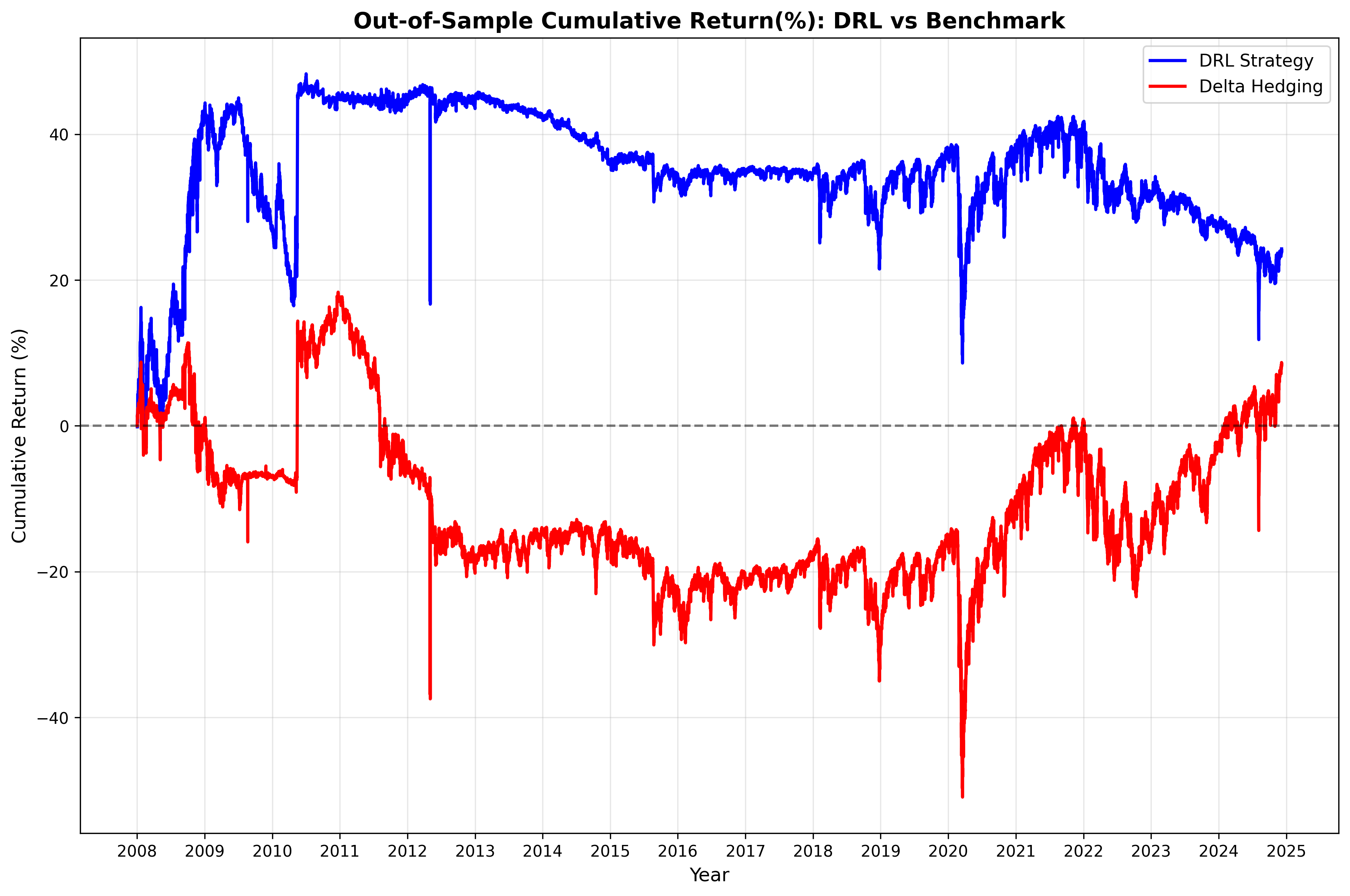}
\caption{Cumulative Return for $C_2$=0.05\%}
\label{fig:cum_pnl_bps5}
\end{figure}

\begin{figure}[H]
\centering
\includegraphics[width=0.8\textwidth]{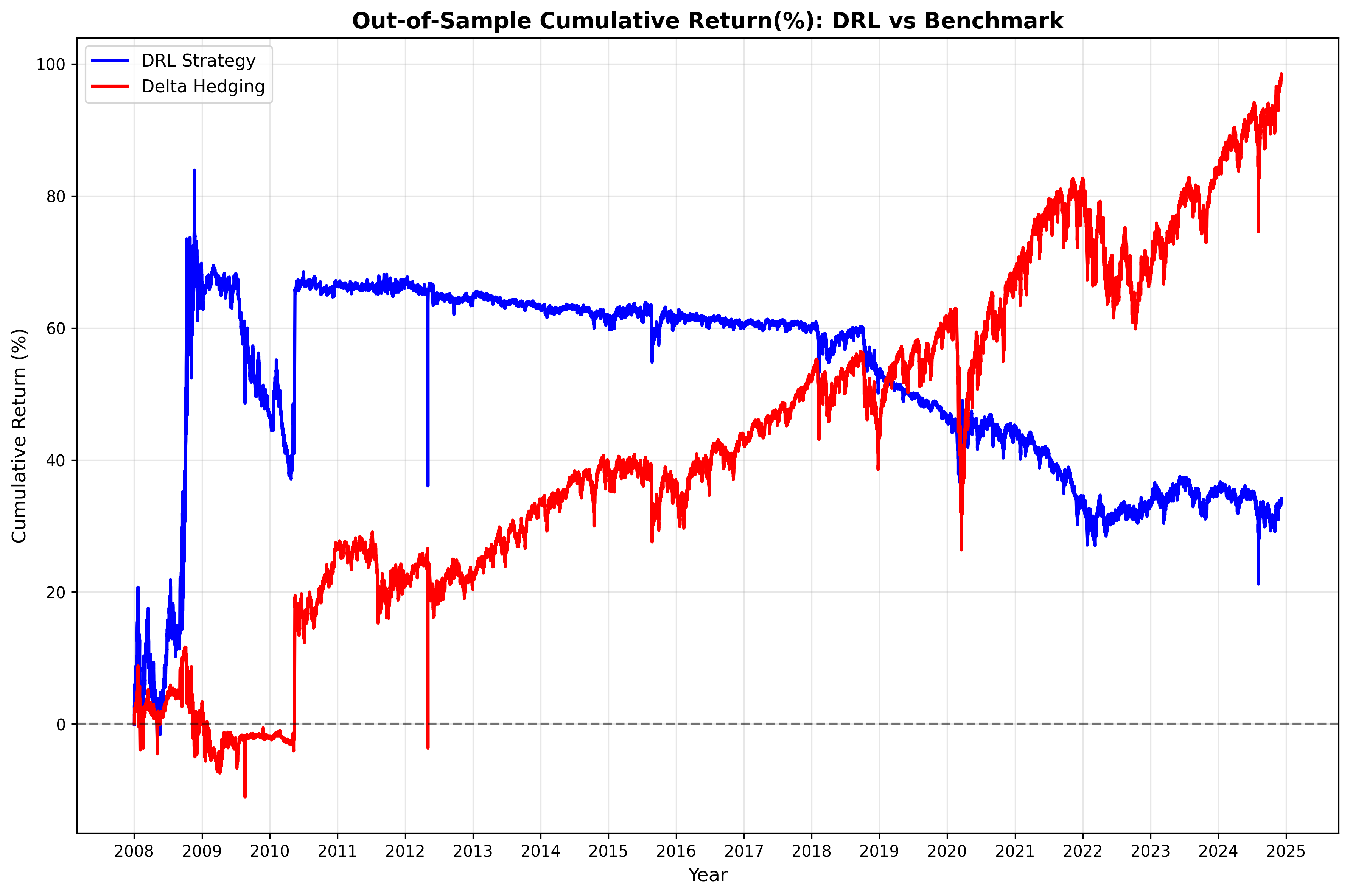}
\caption{Cumulative Return for $C_3$=0.01\%}
\label{fig:cum_pnl_bps1}
\end{figure}

In this part we tried varying the transaction cost parameter, which would penalise frequent large changes in position. The tested parameters were as follows: Scenario 1 - $C_1 = 0.1\%$, Scenario 2 - $C_2 = 0.05\%$ and Scenario 3 - $C_3 = 0.01\%$.\\
Table \ref{tab:drl_vs_delta_3runs} showcases the extended effect of transaction cost on the PnL. While in Scenario 1 both models loose money, the performance significantly improves in the following runs. When the costs are low, the benchmark performs better both in terms of returns or information ratio, but also experiences lower drawdowns. Figures \ref{fig:cum_pnl_bps10}, \ref{fig:cum_pnl_bps5}, and \ref{fig:cum_pnl_bps1} showcase the behaviour of the models across the years.
\begin{table}[H]
\centering
\begin{tabular}{lcccccc}
\hline
{Metric} & \multicolumn{2}{c}{Scenario 1} & \multicolumn{2}{c}{Scenario 2} & \multicolumn{2}{c}{Scenario 3} \\
\cline{2-7}
 & DRL & DH & DRL & DH & DRL & DH \\
\hline
Annualized return (ARC) & -1.99\% & -10.83\% & 1.29\% & 0.52\% & 1.75\% & 4.15\% \\
Annualized Std. dev. (ASD) & 0.0607 & 0.0629 & 0.0591 & 0.0630 & 0.0603 & 0.0629 \\
Sharpe Ratio & -0.3285 & -1.7208 & 0.2185 & 0.0819 & 0.2903 & 0.6592 \\
Information Ratio & 0.0501 & --- & 0.0188 & --- & -0.0909 & --- \\
Adj. Information Ratio & 1.11 & --- & -0.67 & --- & 2.55 & --- \\
Max Drawdown & -74.35\% & -81.90\% & -47.05\% & -59.56\% & -58.96\% & -35.85\% \\
Final PnL (\%) & -28.88\% & -85.64\% & 24.28\% & 9.11\% & 34.17\% & 98.99\% \\
\hline
\end{tabular}

\caption{Comparison of DRL performance vs. delta hedging benchmark across three parameter settings ($C_1$ - $C_3$).}
\label{tab:drl_vs_delta_3runs}
\end{table}


\begin{figure}[H]
\centering
\includegraphics[width=0.8\textwidth]{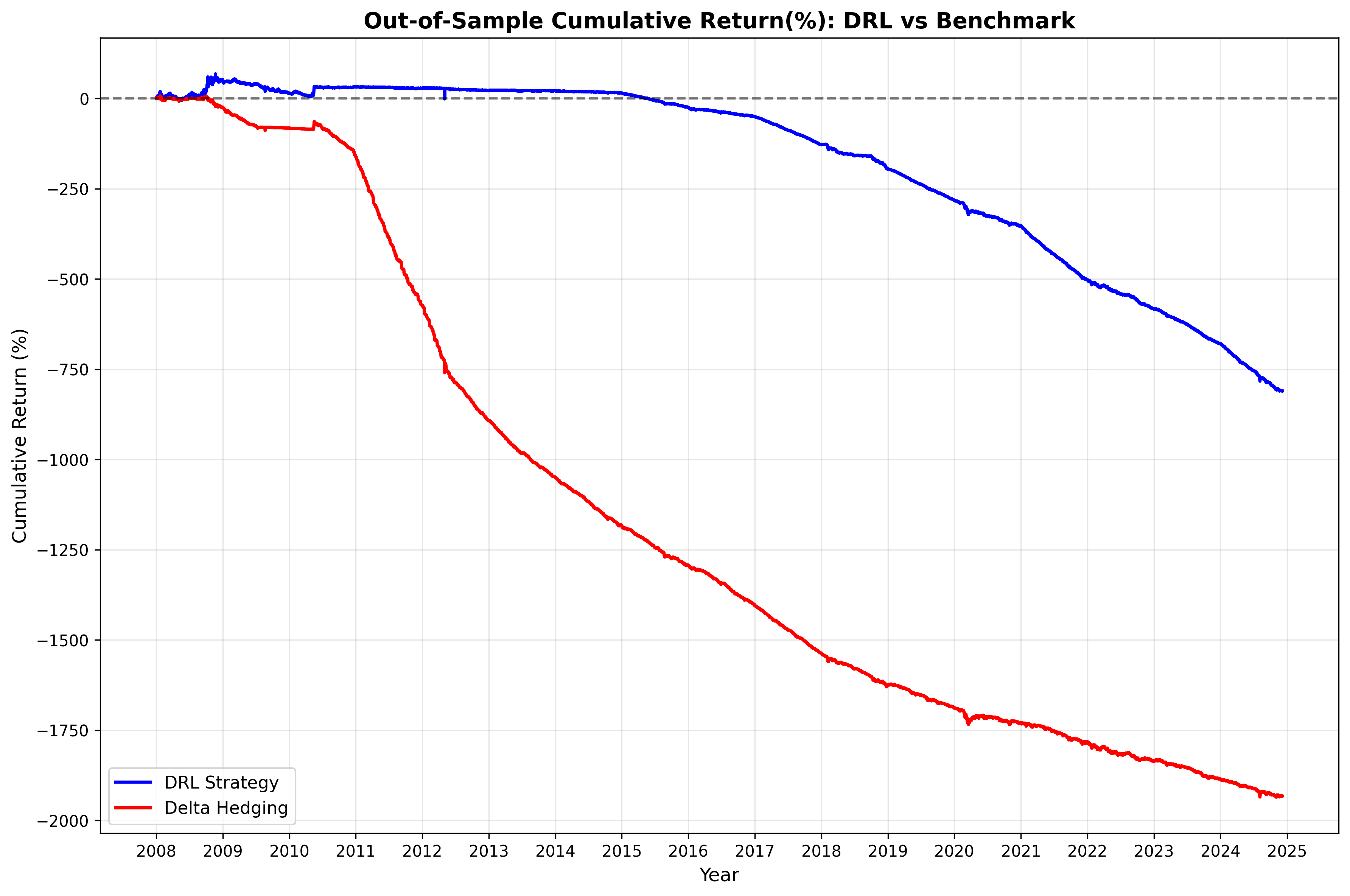}
\caption{Cumulative Return for $C_4$=1\%}
\label{fig:cum_pnl_bps100}
\end{figure}

Finally, we would like to present an extreme case when costs are equal to 1\% Figure \ref{fig:cum_pnl_bps100}. The DRL performs significantly better than benchmark, yet still the agent ends up with huge losses - more than 750\% on the initial investment.
\subsection{Volatility window}
\begin{figure}[H]
  \centering
  \begin{subfigure}{0.48\textwidth}
    \centering
    \includegraphics[width=\linewidth]{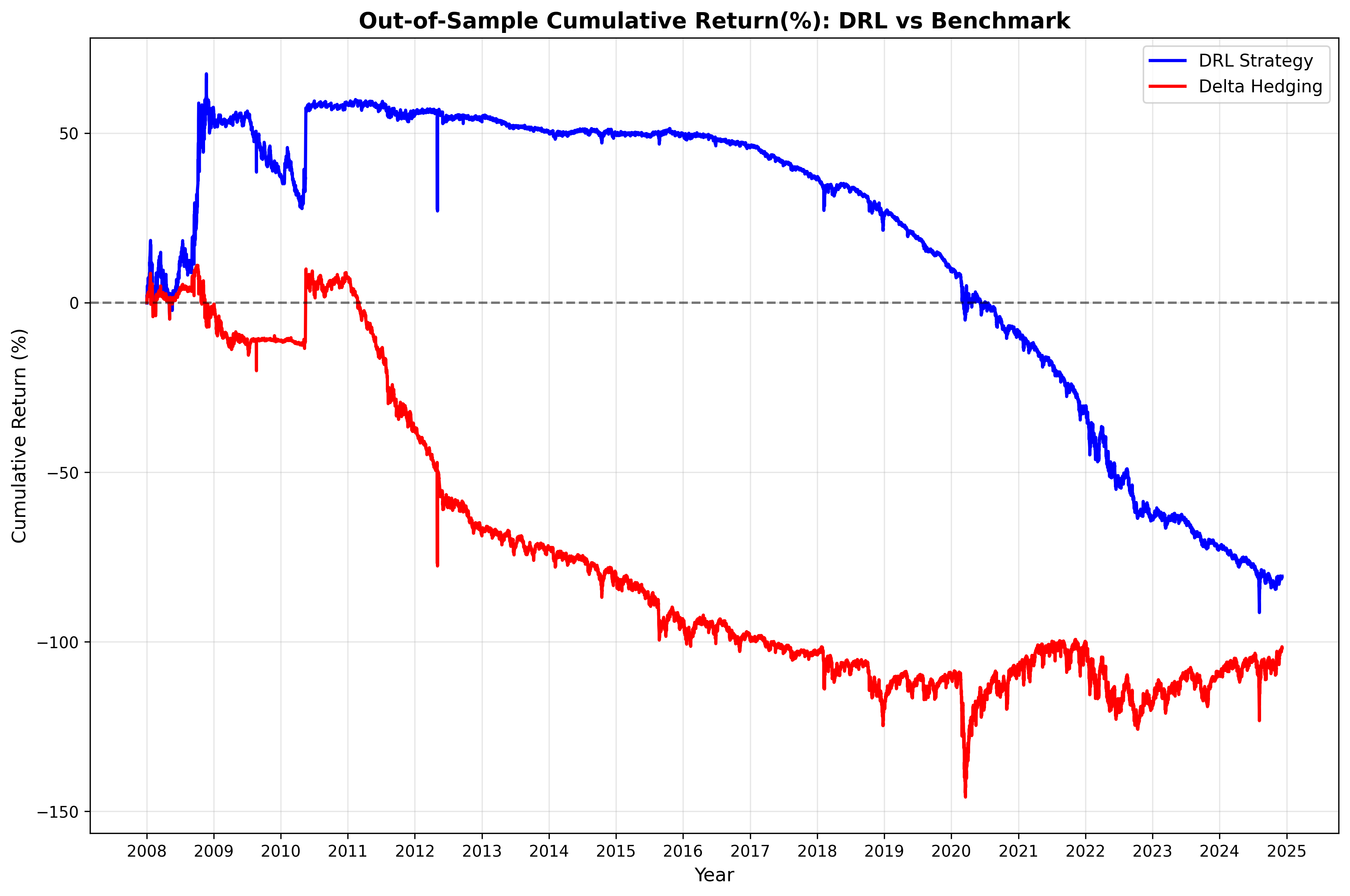}
    \caption{$w = 20$}
    \label{fig:img1}
  \end{subfigure}\hspace{0.03\textwidth}
  \begin{subfigure}{0.48\textwidth}
    \centering
    \includegraphics[width=\linewidth]{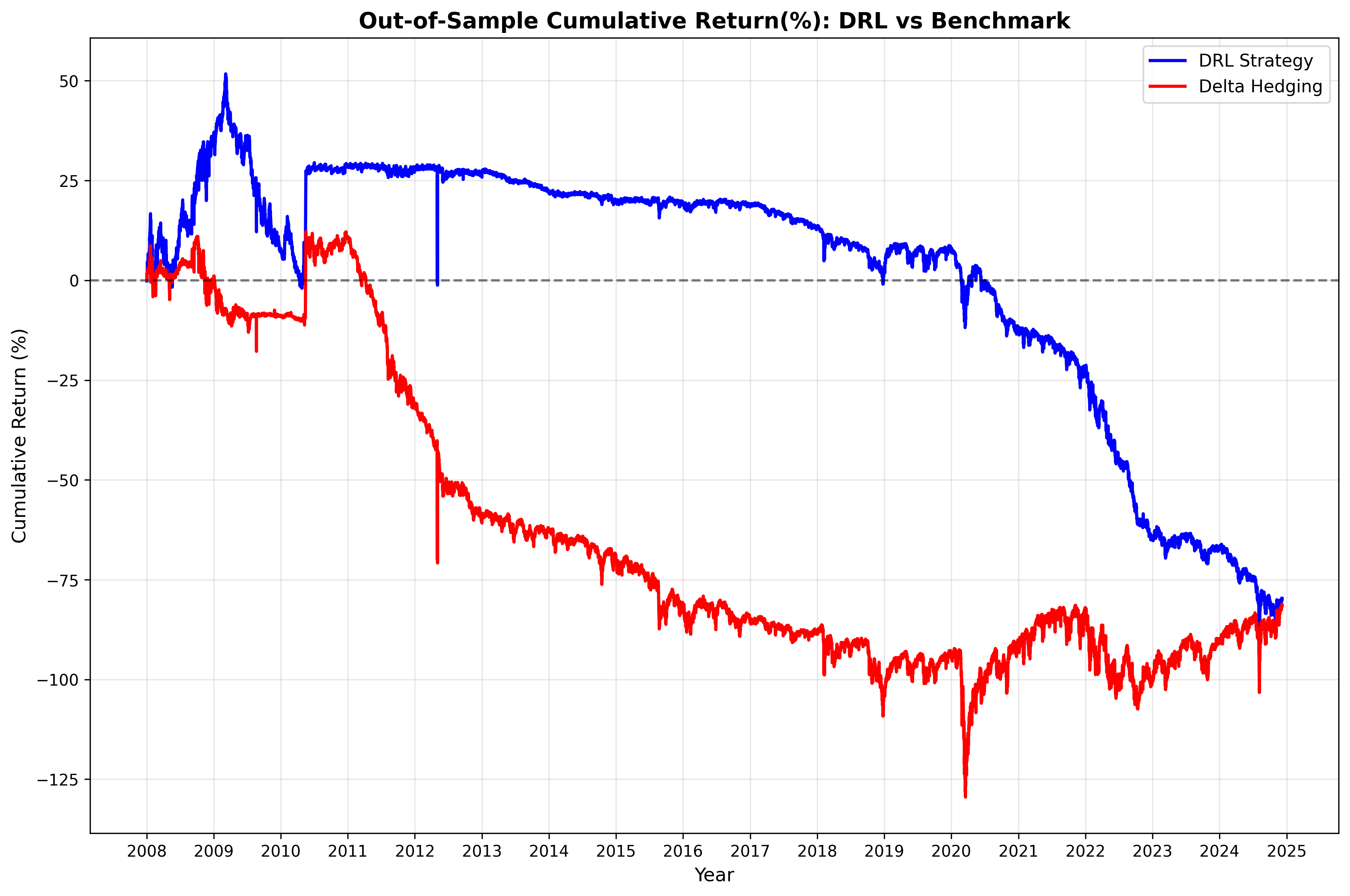}
    \caption{$w = 100$}
    \label{fig:img2}
  \end{subfigure}
  
  \vspace{0.3cm}
  
  \begin{subfigure}{0.48\textwidth}
    \centering
    \includegraphics[width=\linewidth]{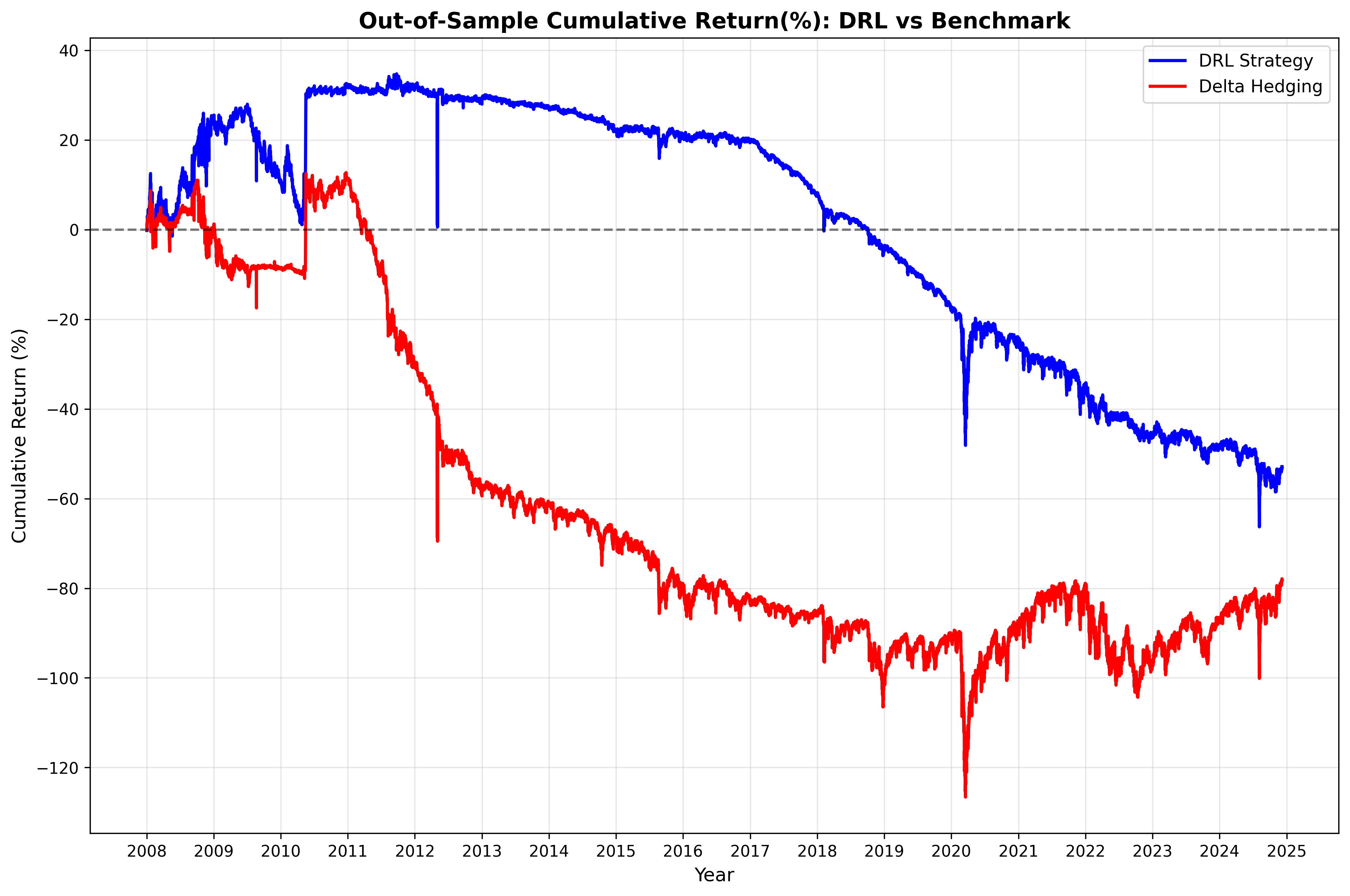}
    \caption{$w = 150$}
    \label{fig:img3}
  \end{subfigure}\hspace{0.03\textwidth}
  \begin{subfigure}{0.48\textwidth}
    \centering
    \includegraphics[width=\linewidth]{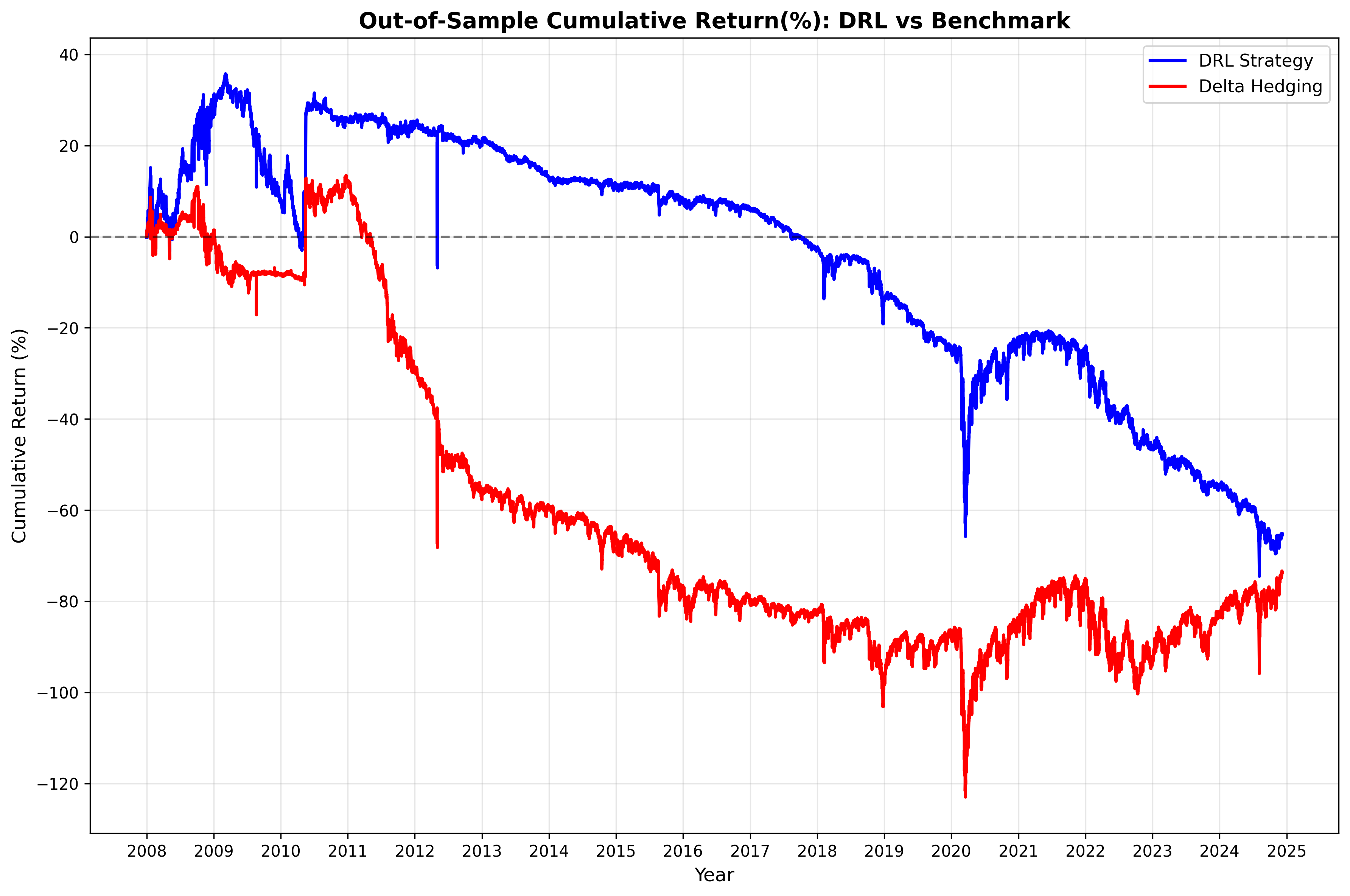}
    \caption{$w = 300$}
    \label{fig:img4}
  \end{subfigure}

  \caption{Cumulative returns for different realized volatility windows (w) - blue represents DRL performance, red - benchmark delta hedging}
  \label{fig:four_images_grid}
\end{figure}

One of the parameters we would like to analyze in more details is a choice of volatility window. When the window is shorter, the model reacts faster to changes in market regimes, which can lead to model taking action on spikes, rather than actual trends. On the other hand, longer window slows down model reaction and can lead to lower robustness. Given the data is in 30-min frequency, we tested windows of length 20, 50, 100, 150 and 300 steps, which correspond to approximately 1.5, 4, 7.5, 11.5, and 23 trading days. The results are in the Table \ref{tab:DRL_volatility} for DRL model and table \ref{tab:dh_volatility} for benchmark.
\begin{table}[H]
\centering
\caption{Performance of DRL strategy across different volatility windows}
\label{tab:DRL_volatility}
\begin{tabular}{lccccc}
\hline
{Metric} & $w=300$ & $w=150$ & $w=100$ & $w=50$ & $w=20$ \\
\hline
Annualized return (ARC) & -0.52\% & -0.43\% & -0.57\% & -0.34\% & -0.58\% \\
Annualized Std. dev. (ASD) & 0.0612 & 0.0567 & 0.0570 & 0.0607 & 0.0579 \\
Sharpe Ratio & -0.412 & -0.429 & -0.450 & -0.385 & -0.446 \\
Information Ratio & 0.000 & 0.029 & -0.014 & 0.050 & 0.005 \\
Adj. Information Ratio & 21.46 & 24.71 & 21.35 & 22.17 & 20.48 \\
Max Drawdown & -76.11\% & -71.32\% & -82.15\% & -74.35\% & -84.39\% \\
Final PnL (\%) & -65.20\% & -52.87\% & -79.69\% & -28.88\% & -80.65\% \\
\hline
\end{tabular}
\end{table}
\begin{table}[H]
\centering
\caption{Performance of Delta Hedging (benchmark) across different volatility windows}
\label{tab:dh_volatility}
\begin{tabular}{lccccc}
\hline
{Metric} & $w=300$ & $w=150$ & $w=100$ & $w=50$ & $w=20$ \\
\hline
Annualized return (ARC) & -0.57\% & -0.59\% & -0.61\% & -0.63\% & -0.71\% \\
Annualized Std. dev. (ASD) & 0.0630 & 0.0630 & 0.0629 & 0.0629 & 0.0630 \\
Sharpe Ratio & -0.408 & -0.412 & -0.415 & -0.418 & -0.430 \\
Information Ratio & --- & --- & --- & --- & --- \\
Adj. Information Ratio & --- & --- & --- & --- & --- \\
Max Drawdown & -79.89\% & -80.59\% & -81.13\% & -81.90\% & -83.98\% \\
Final PnL (\%) & -73.00\% & -77.51\% & -80.64\% & -85.64\% & -101.03\% \\
\hline
\end{tabular}
\end{table}
For delta hedging, the performance clearly deteriorates as the window gets smaller. We observe decrease in annual returns, Sharpe ratio. We also observe higher drawdowns. For DRL, the results are less clear. While we also observe overall trend of decrease in returns and increase in standard deviation or drawdowns, we observe the highest returns for $w = 50$ and lowest drawdowns for $w = 150$. Given the results, we have decided to use $w = 50$ for base case model. Figure \ref{fig:four_images_grid} showcases the models performance across the years.

\subsection{Risk penalty parameter}
\begin{figure}[H]
\centering
\includegraphics[width=0.8\textwidth]{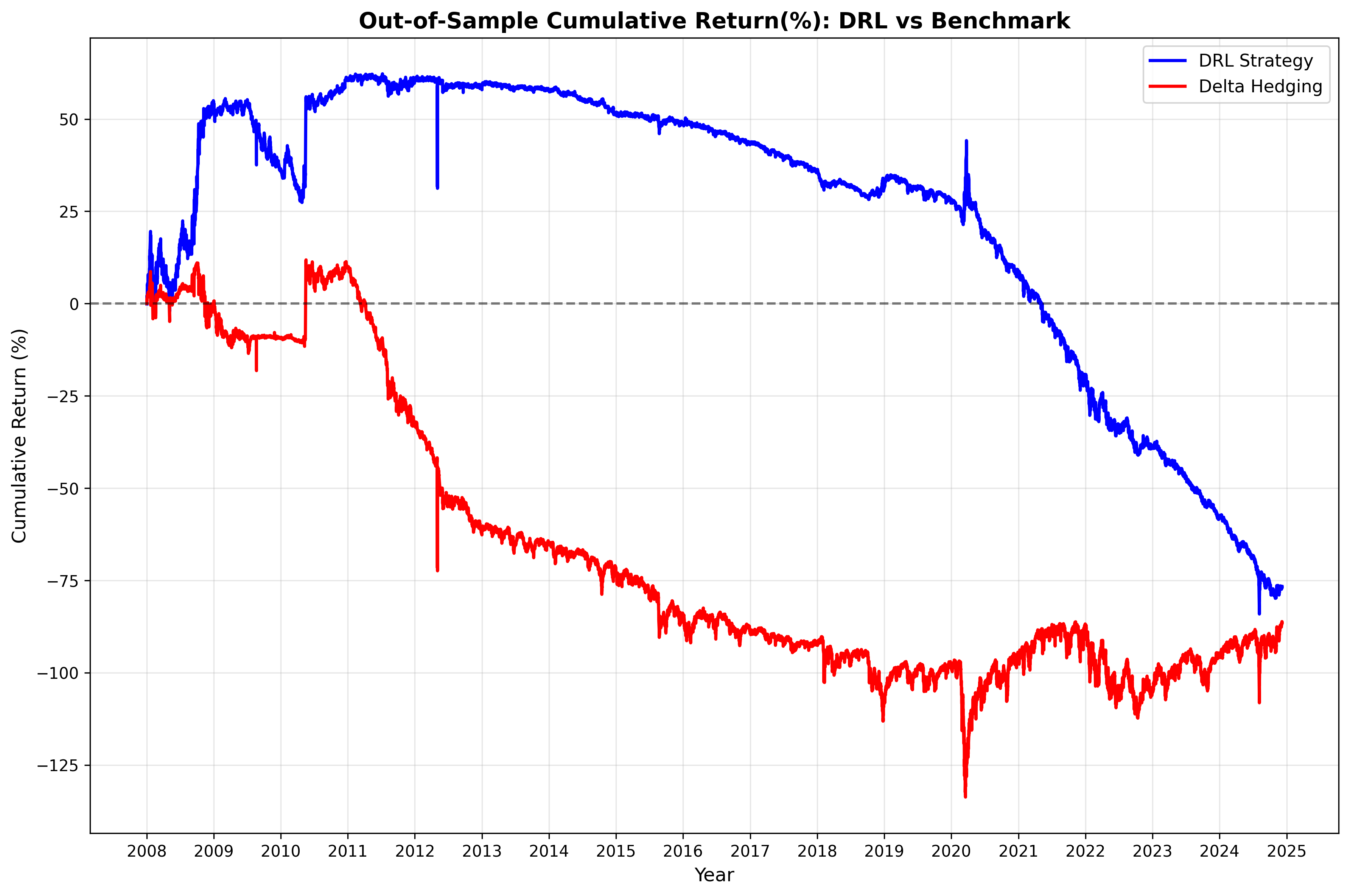}
\caption{Cumulative Return for $\xi$=0.1}
\label{fig:cum_pnl_05}
\end{figure}

\begin{figure}[H]
\centering
\includegraphics[width=0.8\textwidth]{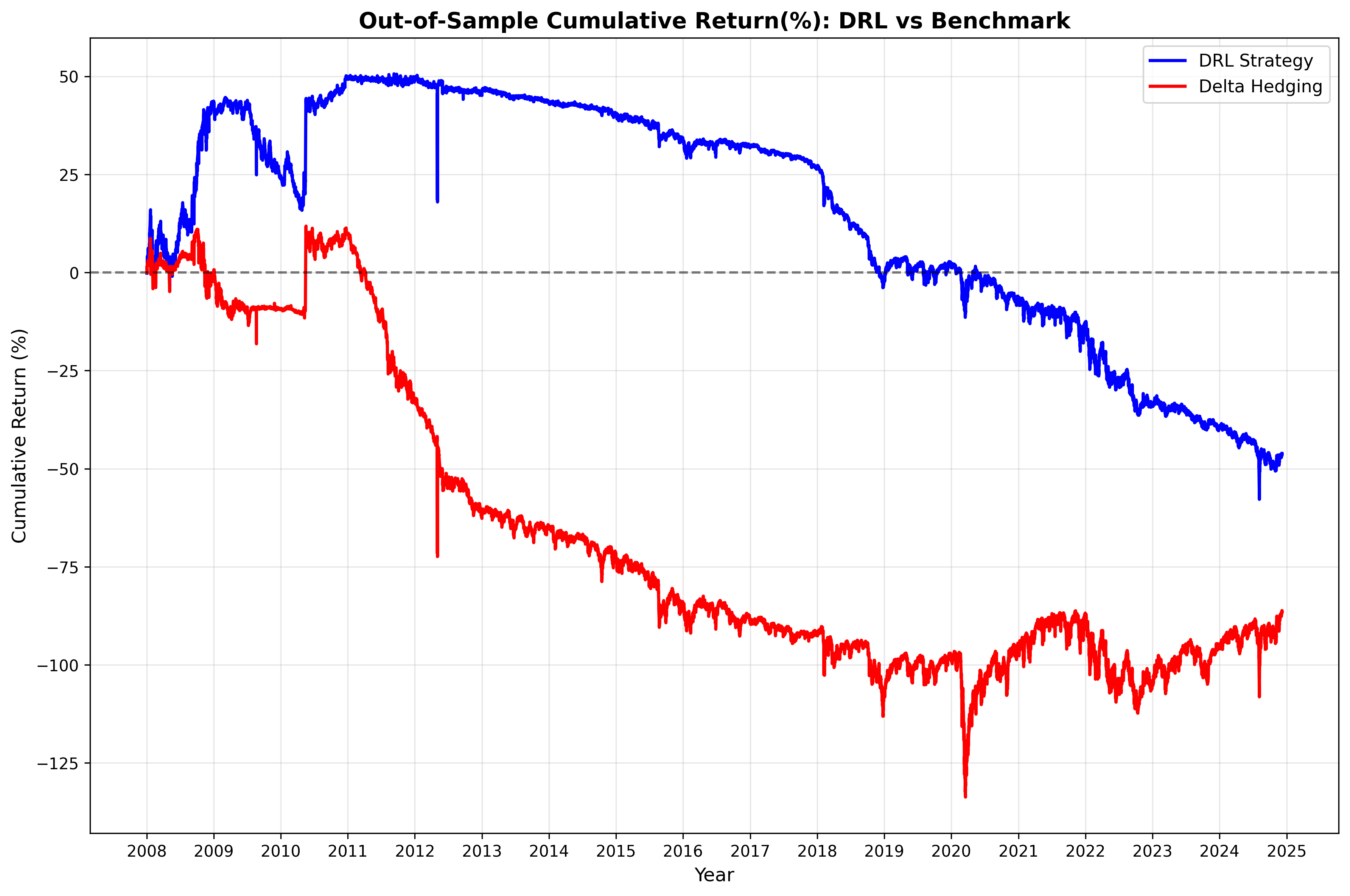}
\caption{Cumulative Return for $\xi$=0.005}
\label{fig:cum_pnl_005}
\end{figure}

\begin{figure}[H]
\centering
\includegraphics[width=0.8\textwidth]{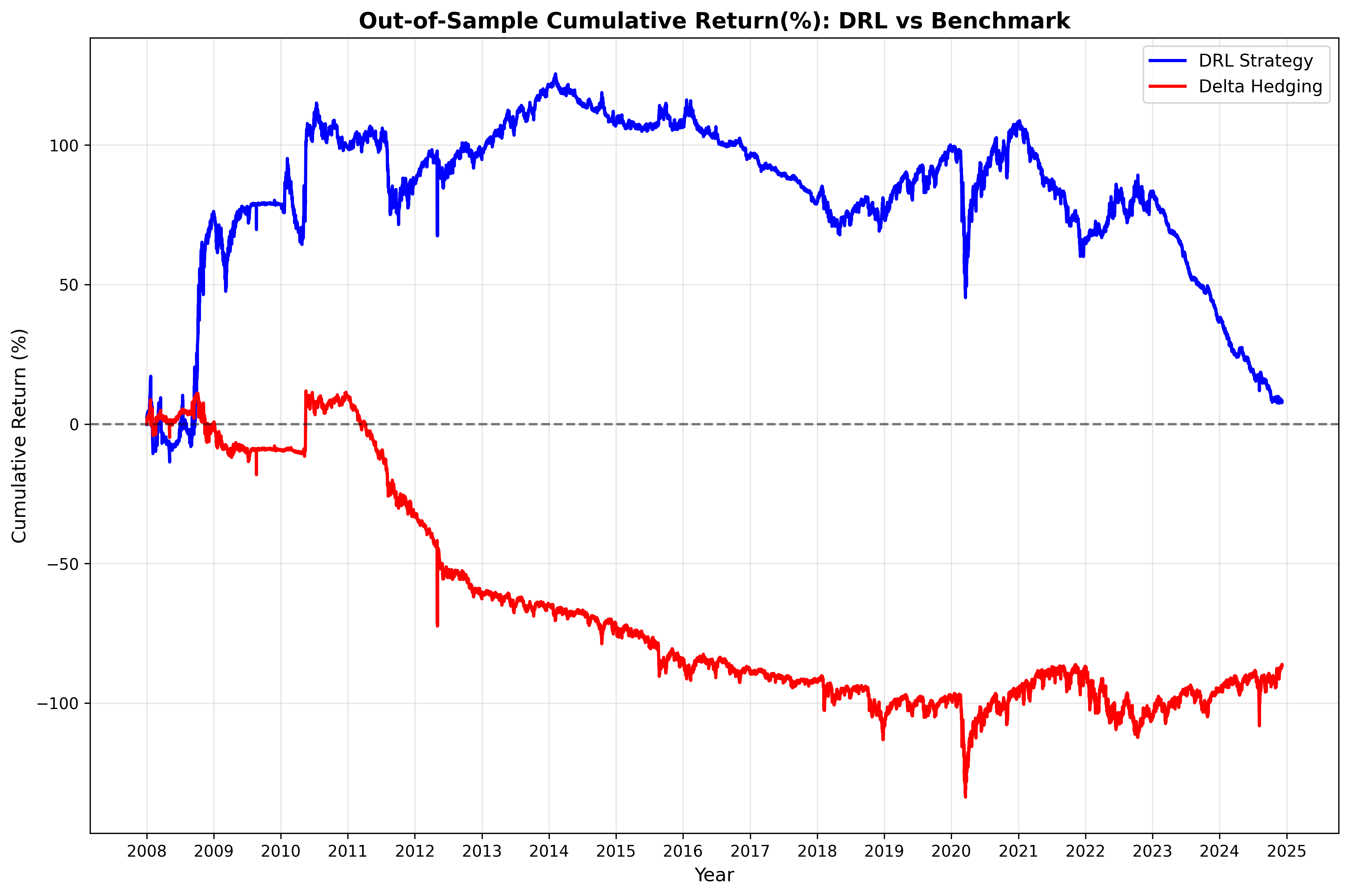}
\caption{Cumulative Return for $\xi$=0.001}
\label{fig:cum_pnl_001}
\end{figure}

At first we have tried varying the penalty parameter $\xi$ in a reward function. We tried values of 0.1, 0.005, 0.001 and additionally 0.01 in base case. We have observed that with the higher risk penalty parameter, the performance of the DRL agent was worse (Figures \ref{fig:cum_pnl_005}, \ref{fig:cum_pnl} \ref{fig:cum_pnl_05}. During all those runs trading costs were set to 0.1\%.
\begin{table}[H]
\centering
\caption{Comparison of DRL strategy vs. delta hedging benchmark across three different risk penalty values $\xi_1 = 0.001$, $\xi_2 = 0.005$, $\xi_3 = 0.1$}
\label{tab:drl_vs_delta_runs2}
\begin{tabular}{lcccccc}
\hline
{Metric} & \multicolumn{2}{c}{Scenario 1} & \multicolumn{2}{c}{Scenario 2} & \multicolumn{2}{c}{Scenario 3} \\
\cline{2-7}
 & DRL & DH & DRL & DH & DRL & DH \\
\hline
Annualized return (ARC) & -0.54\% & -0.63\% & -0.39\% & -0.63\% & -0.24\% & -0.63\% \\
Annualized Std. dev. (ASD) & 0.0553 & 0.0629 & 0.0549 & 0.0629 & 0.0738 & 0.0629 \\
Sharpe Ratio & -0.460 & -0.418 & -0.435 & -0.418 & -0.304 & -0.418 \\
Information Ratio & -0.0108 & --- & 0.0527 & --- & 0.0541 & --- \\
Adj. Information Ratio & 0.17 & --- & -0.05 & --- & -0.05 & --- \\
Max Drawdown & -81.41\% & -81.90\% & -72.86\% & -81.90\% & -76.93\% & -81.90\% \\
Final PnL (\%) & -76.66\% & -85.64\% & -46.12\% & -85.64\% & 8.19\% & -85.64\% \\
\hline
\end{tabular}
\end{table}
The results in table \ref{tab:drl_vs_delta_runs2} indicate that both models on average loose money. Both models also experience very significant drawdowns. Nevertheless, the agent managed to have multiple positive PnL periods, and overall has positive Information ratio. Despite negative returns, we can say that DRL outperformed the benchmark. Figures \ref{fig:cum_pnl_05}, \ref{fig:cum_pnl_005}, and \ref{fig:cum_pnl_001} present graphically results for different values of $\xi$. We can clearly see the better performance of the DRL model with smaller risk penalty. The deterioration is the most visible in the last few year of the out-of-sample period, when the volatility of prices was significant.
The final observation concluding this sensitivity analysis is that while all parameters for which we performed sensitivity analysis do impact the models' performance, the impact trading costs level have is far more significant, mostly due to benchmark performance drastically decreasing. 
\section{Conclusions}
The goal of this project was to showcase whether the deep reinforcement learning model trained on single historical continuous time series of option prices can outperform a Black-Scholes delta hedging strategy. In order to do this, we gathered data of option prices on S\&P 500 index from years 2004-2024 and and at each time step selected at-the-money option with maturity close to 30 days. We implemented Twin Delayed Deep Deterministic Policy Gradient framework using Pytorch and trained the model with walk-forward approach. On the 17 years long out-of-sample period the DRL agent and delta hedging performance was evaluated using cumulative PnL and additional performance metrics such as information ratio, max drawdown or Sharpe ratio.\par
The general observation from the study is that the DRL model outperforms the delta hedging strategy and achieves higher returns in all but one scenarios. The agent on average had higher Sharpe ratio and lower max drawdown. The deep hedging framework's advantage is particularly evident in high-cost environments, when the agent tends to adapt to market conditions, while the benchmark’s performance deteriorates sharply. However, we observe that the DRL model is also very sensitive to environment parameters, such as volatility, transaction costs and risk-awareness. During the sensitivity analysis, we observed that the increase in risk aversion reduces returns, while shorter volatility estimation intervals make the model more responsive to jumps and also lower performance. Similarly, higher transaction cost parameter decrease the DRL model's performance. This issue could be potentially mitigated by less frequent rebalancing, as current setup of trading every 30 minutes drastically increase the costs. Moreover, we would like to highlight that the number of different model parameters to be tuned makes it difficult to reach the full potential of the DRL framework. In this paper, we have made some assumptions about the methodology such as network architecture or PnL and reward accounting method. While this study adds new perspective to deep hedging, by introducing an agent trained on single time series, we believe that further research needs to be done to examine the full range of opportunities of applying deep reinforcement learning to options hedging.\par
The analysis conducted in this paper allows us to affirmatively answer the research question. The DRL model outperforms traditional delta hedging. Our results showcase sufficient robustness of the framework. Long out-of-sample period and frequency of trading further affirms our conclusion on adaptability and superiority, despite the significant drawdowns of the DRL strategy. \par
This study presented successful implementation of a deep hedging framework to historical S\&P 500 data. We evaluated the results in different market conditions across long period of time. The further research could involve extended empirical experiments with multiple instruments in a portfolio, employing a risk measure in the objective function or extension of the methodology to other markets or derivative types. To extend the existing methodology, further research could include analysing different predictor variables or evaluating different rebalancing frequencies.

\section*{Source code}
The complete code can be found and downloaded from Github repository:\\ 
https://github.com/zof-br/DRL\_hedging
\newpage
\bibliography{references}
\end{document}